\documentclass[aps,prd,reprint,superscriptaddress,floatfix]{revtex4-1}

\usepackage[utf8x]{inputenc}
\usepackage{microtype}

\usepackage{graphicx}
\usepackage[dvipsnames]{xcolor}
\usepackage{rotating}

\usepackage{amsmath}
\usepackage{amssymb}
\usepackage{amsfonts}

\usepackage{tabu}
\usepackage{booktabs}
\usepackage{multirow}
\usepackage{array}
\usepackage{enumitem}
\usepackage{url}
\usepackage[colorlinks=true,urlcolor=blue,linkcolor=blue,citecolor=blue]{hyperref}

\newcommand{\abs}[1]{\ensuremath{\left\vert#1\right\vert}}

\def\0#1#2{\frac{#1}{#2}}

\def\s0#1#2{\mbox{\small{$ \frac{#1}{#2} $}}}

\begin{document}

\title{Quantum Improved Schwarzschild-(A)dS and Kerr-(A)dS Space-times}

\author{Jan M.~Pawlowski}
\affiliation{Institut für Theoretische Physik, Universit\"at Heidelberg,
Philosophenweg 16, 69120 Heidelberg, Germany}
\affiliation{ExtreMe Matter Institute EMMI, GSI Helmholtzzentrum für
Schwerionenforschung mbH, Planckstr.\ 1, 64291 Darmstadt, Germany}
\author{Dennis Stock}
\affiliation{University of Bremen, Center of Applied Space Technology and 
	Microgravity (ZARM), 28359 Bremen, Germany}
\affiliation{Institut für Theoretische Physik, Universit\"at Heidelberg,
	Philosophenweg 16, 69120 Heidelberg, Germany}

\begin{abstract}
  We discuss quantum black holes in asymptotically safe quantum
  gravity with a scale identification based on the Kretschmann
  scalar. After comparing this scenario with other scale
  identifications, we investigate in detail the Kerr-(A)dS and
  Schwarzschild-(A)dS space-times. The global structure of these
  geometries is studied as well as the central curvature singularity
  and test particle trajectories. The existence of a Planck-sized,
  extremal, zero temperature black hole remnant guarantees a stable
  endpoint of the evaporation process via Hawking radiation.

\end{abstract}

\maketitle

\section{Introduction}\label{sec:intro}

The consistent quantisation of gravity is an open challenge to
date. One of the candidates is the asymptotic safety (AS) scenario for
quantum gravity \cite{Weinberg:1980gg}, its attraction being the
possible quantum-field theoretical ultraviolet completion of the Standard
Model with gravity.  If realised, it is the minimal UV closure of high
energy physics including gravity within a purely field-theoretical
set-up.

One of the prominent and characteristic properties of asymptotically
safe gravity is its ultraviolet scaling regime for momentum scales $k$
larger than the Planck scale $M_{\rm pl}$. In the AS set-up, the
latter is defined as the scale beyond which quantum gravity
corrections dominate the physics and agrees well with the classical
Planck scale. In this regime, the Newton's coupling $G$ and
cosmological constant $\Lambda$, as well as all further couplings of
terms, e.g.\ of the higher curvature invariants $R^n$, run according
to their canonical scaling.  For the Newton's coupling and
cosmological constant in particular, this entails
$G(k) \propto 1/{k^2}$ and $\Lambda(k) \propto k^2$ respectively,
instead of the classical constant behaviour. Consequently, the physics
at these scales looks rather different to that of general relativity.

Black holes offer one of the few possibilities where such deviations
from classical general relativity may be observed as they feature
large curvatures. Asymptotically safe quantum black holes have been
amongst the first applications of asymptotically safe gravity after
its first explicit realisation within the functional renormalisation
group \cite{Reuter:1996cp}. Within such a renormalisation group
setting, the Newton's coupling and cosmological constant are naturally
elevated to couplings running with the momentum (RG) scale $k$. Then,
classical solutions of the Einstein field equations are quantum
improved by replacing Newton's and the cosmological constant by
functions depending on a respective length scale. The $k$-dependent
RG-runnings, equipped with an identification between momentum and
length scales, serve as an ansatz for these functions. The earliest
works investigated the Schwarzschild space-time
\cite{Bonanno:1998ye,Bonanno:2000ep} followed by studies of the Kerr
space-time \cite{Reuter:2010xb} and Schwarzschild-(A)dS geometries
\cite{Koch:2013owa}. Black holes in higher dimensions have been
studied in \cite{Falls:2010he}. All works, summarised in
\cite{Koch:2014cqa}, match the classical results of general relativity
in the low energy limit, but show significant changes for the number
of horizons, test particle trajectories, the Hawking temperature, and
the entropy around the Planckian regime.  There is evidence for a
cold, extremal Planck-sized remnant, which is a smallest black hole
with zero temperature, a possibly promising answer to the endpoint of
black hole evaporation. By studying dynamical, non-vacuum solutions
such as the Vaidya space-time, the processes of black hole formation
\cite{Bonanno:2016dyv} and evaporation \cite{Bonanno:2006eu} can be
addressed directly, leading to the same conclusions as above. The
quantum effects render the central curvature singularity at $r=0$ less
divergent, some scenarios lead to a complete resolution. A detailed
study on the implications for the laws of black hole thermodynamics
was performed in \cite{Falls:2012nd}. Most of the above results for a
quantum improved space-time were obtained by using a cut-off
identification based on a classical space-time. This was addressed in
\cite{Reuter:2004nv} and \cite{Emoto:2005te}, where a consistent
framework with an underlying quantum space-time was introduced.

In this work, we present a new scale identification based on the
quantum improved classical Kretschmann scalar. This approach takes
the running of the couplings into account which removes unphysical
features in the resulting geometries. For the first time in this
quantum gravity set-up, the Kerr-(A)dS geometry, as the most general
vacuum black hole solution including a cosmological constant, is
studied in great detail. As a special case ($a=0$), the results for
Schwarzschild-(A)dS are presented separately. The ordinary
Schwarzschild and Kerr solutions are also contained by setting the
cosmological constant to zero.

This work is structured as follows: we start with a brief review of
the AS scenario of quantum gravity in \autoref{sec:ASQG}, and discuss
the studied geometries in \autoref{sec:geometries}. The novel scale
identification is discussed in \autoref{sec:scale}. Results on
horizons and the GR-limit are presented in \autoref{sec:horizons}, the
global structure in \autoref{sec:penrose}, test particle trajectories
in \autoref{sec:trajectories}, the curvature singularity in
\autoref{sec:singularity}, and Hawking temperatures and the black hole
evaporation process in \autoref{sec:temperature}. Some technical
details are deferred to the appendices which contain in particular a
discussion of proper distance matchings, see
appendix \ref{app:othermatchings}.

\section{Asymptotic Safe Quantum Gravity}
\label{sec:ASQG}
By now, asymptotically safe quantum gravity has been studied in
  an impressive wealth and depth of approximations including higher
  derivative terms, the full $f(R)$ potential as well as the 
inclusion of matter, see
  e.g.\
  \cite{Niedermaier:2006ns,Litim:2011cp,Reuter:2012id,Bonanno:2017pkg,
    Percacci:2017fkn,Eichhorn:2017egq} and references therein. The specific shape of the
  running of $G(k)$ and $\Lambda(k)$ depends on the regularisation
  scheme or regulator which also defines part of the scale
  identification. Moreover, despite the advances in the approximation
  schemes used in recent computations, the systematic error estimates 
are still relatively large. However, while these details do not affect the
  results of this work qualitatively, all runnings have to meet the
  following general constraints: 
\begin{enumerate}
\item The existence of a UV fixed point, that is, the dimensionless
  couplings $g$ and $\lambda$ become constant in the UV-limit:
	\begin{equation}
	(g, \lambda)\overset{k\rightarrow\infty}{\longrightarrow}(g_*,\lambda_*)\;.
	\end{equation}
\item The effective theory should recover the classical theory
        of general relativity in the IR-limit, i.e. $G$ and $\Lambda$
        approach Newton's constant $G_0$ and a cosmological constant $\Lambda_0$
        respectively, reducing the effective action to the
        Einstein-Hilbert action:
	\begin{equation}
   \hspace{1cm}       G,\Lambda\overset{k\rightarrow0}{\longrightarrow} G_0,
          \Lambda_0\quad\Leftrightarrow\quad g\sim k^2\ \ \text{and}\ \  \lambda\sim k^{-2}\;.
	\label{irbehaviour}
	\end{equation}
\end{enumerate}
The running of $g(k)$ and $\lambda(k)$ is typically obtained
numerically. In the following, we approximate them by analytical
expressions, which show the same features and are compatible with the
above constraints in the UV and IR. For instance, a comparison with
the results of the systematic vertex expansion up to the fourth order
in \cite{Denz:2016qks} is provided in \autoref{fig:vertexexp} in the
appendix. The following scale runnings are used, 
\begin{align}
  g(k)&=\frac{G_0 g_* k^2}{g_*+G_0 k^2}&\quad&\Leftrightarrow\quad G(k)=
  &\!\!\!\!\!\!\frac{G_0 g_*}{g_*+G_0 k^2}\;,\nonumber\\[2ex]
  \lambda(k)&=\frac{\Lambda_0}{k^2}+\lambda_*&\quad&\Leftrightarrow
                                                     \quad\Lambda(k)=&\!\!\!\!\!\!
                                                                       \Lambda_0+\lambda_* k^2\;.
                                                                       \label{analyticlambda}
\end{align}
The functional dependence of $g(k)$ was already used in \cite{Bonanno:2000ep} and $\lambda(k)$ agrees with the expression used in \cite{Koch:2014cqa} without the logarithmic term. $G_0$ and $\Lambda_0$ are
the IR-values of the gravitational and cosmological coupling, whereas
$g_*$ and $\lambda_*$ are the fixed point values of the dimensionless
couplings. In the following analysis, we choose the numerical values
at the fixed point to be the ones for the background couplings obtained in appendix B of
\cite{Denz:2016qks}, together with their identification scheme in (34): 
\begin{align}
 g_*=1.4\,,\qquad\qquad  \lambda_*=0.1\,.
\end{align}
The dependence of Newton's coupling $G$ and cosmological constant
$\Lambda(k)$ on the running scale $k$ reflects the non-trivial
dependence of the full effective action at vanishing cut-off scale on
the Laplacian $\Delta$, as well as the existence of higher order
terms. As in earlier works, we use the following strategy to take into
account these terms: we use solutions to the Einstein field equations
and assume that quantum gravity effects can be modeled by
momentum-dependent $G$ and $\Lambda$, equipped with a relation to
convert the momentum into a length scale. The now $r$-dependent $G$
and $\Lambda$ are inserted back into the classical solution, yielding
a quantum improved space-time. This procedure is the analogue of the
Uehling's correction in QED, see \cite{RGimprovement} and
\cite{Bonanno:2000ep} for more details. In the context of
asymptotically save gravity, it has been shown in \cite{Koch:2013owa},
that a quantum improved metric in the above sense can be a solution to
the field equations derived from the quantum improved Einstein-Hilbert
action in the UV-limit, at least in the spherically symmetric
case. Furthermore, the quantum improved metric, together with its
observables, approach the results obtained from general relativity in
the IR, and thus show the correct low energy limit.

In the following we need the couplings $G(r)$ and $\Lambda(r)$ as
functions of radius $r$ rather than momentum scale $k$. Thus, we have
to establish a relation $k(r)$ in order to arrive at
$G(k(r))\,,\, \Lambda(k(r))$.  A commonly used ansatz for $k(r)$ is
\begin{equation}
k(r)= \frac{\xi}{D(r)}\;,
\label{genscale}
\end{equation}
with constant $\xi$ and a $r$-dependent function $D$ with momentum 
dimension minus one (length), encoding the physical scales. Our choice
$\xi=1/\sqrt{\lambda_*}$ is further motivated in appendix \ref{app:choiceofxi}.

\section{Investigated Geometries}
\label{sec:geometries}
In this work, we study geometries based on solutions of the Einstein
equations with cosmological constant, but vanishing
stress-energy tensor. Depending on the sign of the cosmological
constant, the space-time is called asymptotically de Sitter (dS),
flat, or anti-de Sitter (AdS). As the stress-energy tensor is
zero, the black hole is allowed to have a mass and angular momentum,
but no charge. Thus, we study the Schwarzschild-(A)dS space-time of a
non-rotating black hole and the Kerr-(A)dS space-time for a rotating
black hole. 

The Kerr-(A)dS geometry is the most general vacuum black hole
solution, which includes a cosmological constant. Hence the
Schwarzschild-(A)dS as well as the Schwarzschild and Kerr solutions in
flat space can be obtained from Kerr-(A)dS by either setting the
rotations parameter $a$ or the cosmological coupling $\Lambda$ to
zero. In our analysis, we discuss the quantum improved
Schwarzschild-(A)dS and Kerr-(A)dS solution, but the results can be
easily extended to asymptotically flat space-times. Below we briefly summarise 
some basic properties of these geometries.

\subsection{Schwarzschild-(A)dS}
The Schwarzschild-(A)dS solution is a two-parameter family of
solutions of the non-vacuum Einstein equations, labeled by
$(M,\Lambda)$. It is explicitly given by
\begin{align}
\mathrm{d}s^2 &= -f(r) \mathrm{d}t^2 +f^{-1}(r)\mathrm{d}r^2 + 
r^2\mathrm{d}\Omega^2\;,\nonumber\\[2ex]
f(r)&:=1-\frac{2MG}{r}-\frac{\Lambda}{3}r^2\;,
\label{schwads}\end{align}
with $t\in(-\infty,\infty)$, $r\in(0,\infty)$, Newtons's constant $G$,
the cosmological constant $\Lambda$ and $\mathrm{d}\Omega^2$ the
metric on $S^2$. This solution is spherically symmetric and displays a curvature singularity at $r=0$ if $M\neq0$. For
$\Lambda=0$, it reduces to the Schwarzschild solution in flat space
and for $M=0$ but $\Lambda\neq0$, one obtains the metric describing
AdS or dS, depending on the sign of $\Lambda$.  Therefore, this metric
interpolates between a Schwarzschild solution on small scales and an
(A)dS solution on large scales. Horizons are solutions to $f(r)=0$.
\subsection{Kerr-(A)dS}
The Kerr-(A)dS solutions form a three parameter family, labelled by
($M, J, \Lambda$). Unlike in the flat case, $M$ and $J$ cannot be
interpreted as mass and angular momentum of the black hole anymore,
however, for convenience we still refer to them as mass and angular
momentum in the text below. The metric is given by
\cite{Gibbons:1977mu}, 
\begin{align}
\mathrm{d}s^2 = &-\frac{\Delta_r}{\rho^2 \Xi^2}\left(\mathrm{d}t-a\sin^2\theta 
\mathrm{d}\phi\right)^2+\frac{\rho^2}{\Delta_r}\mathrm{d}r^2+
\frac{\rho^2}{\Delta_\theta}\mathrm{d}\theta^2\nonumber\\[2ex]
&+\frac{\Delta_\theta
 \sin^2\theta}{\Xi^2\rho^2}
\left(a \mathrm{d}t-(r^2+a^2)\mathrm{d}\phi\right)^2\;,
\label{kerrads}
\end{align}
with 
\begin{align}
a&:= \frac{J}{M}\;,\nonumber\\[2ex]
\rho^2&:=r^2+a^2\cos^2\theta\;,\nonumber\\[2ex]
\Delta_r&:=(r^2+a^2)(1-\frac{\Lambda}{3}r^2)-2G M r\;,\nonumber \\[2ex]
\Delta_\theta&:=1+\frac{\Lambda}{3}a^2\cos^2\theta\;,\nonumber\\[2ex]
\Xi&:=1+\frac{\Lambda}{3}a^2\;.
\label{definitions}
\end{align}
The parameter $a$ is referred to as rotation parameter and is restricted by 
\begin{equation}
\frac{1}{3}\Lambda a^2>-1 \;, 
\label{a_restriction}
\end{equation}
in order to preserve the Lorentzian signature of the metric. The
coordinate ranges are $t\in(-\infty,\infty)$, $r\in(0,\infty)$,
$\theta\in[0,\pi]$ and $\phi\in[0,2\pi)$.  It can be shown that this
solution reduces to a Kerr black hole in the limit of small $r$,
whereas for large $r$ it gives back the metric of (A)dS. In the case
of $a=0$, one recovers the Schwarzschild-(A)dS metric of a
non-rotating black hole (\ref{schwads}). For $\Lambda=0$, the metric
reduces to the one of a Kerr black hole in flat space. For $M=0$ and
$a=0$, we recover (A)dS. For $M\neq0$, there is a curvature
singularity at $r=0$ in the equatorial plane
$\theta=\frac{\pi}{2}$. Horizons correspond to solutions of
$\Delta_r=0$.

\section{Scale Identification}\label{sec:scale}
In pure gravity systems, i.e. systems with vanishing stress-energy
tensor, there is no unique way to fix the scale identification. In
fact, it turns out that physical features of the space-time such as
the number of horizons, Hawking temperatures and the strength of the
curvature singularity actually do depend on the particular choice of
$k(r)$. Motivated by dimensional analysis, one simple way to identify
the momentum scale of the FRG set-up with a length scale is an inverse
proportionality. However, this ansatz is completely insensitive to
typical scales of the underlying space-time. Therefore, different
scale setting procedures have been brought forward, for instance on
the level of the field equations, e.g.\ \cite{Koch:2013owa}. A more
feasible approach to account for space-time features is to use proper
distance integrals. As such, they give rise to diffeomorphism
invariant quantities. Proper distance integrals based on classical
space-times were suggested in \cite{Bonanno:2000ep}. Later, it was
pointed out in \cite{Reuter:2004nv, Emoto:2005te}, that this procedure
can be upgraded to a consistent setting by computing the proper
distance already in the quantum improved geometry.

Here, we investigate this approach for Schwarzschild-(A)dS and
Kerr-(A)dS space-times. However, using two different integrations
contours for the computation of the proper distance in the upgraded
scheme yields ill-defined quantities. In case of a radial integration
path, we find diverging surface gravities for all horizons. This results 
in divergent Hawking temperatures, independent of the black hole
parameters. In case of a path prescribed by the timelike geodesic of
an infalling observer, we find an identically vanishing
eigentime. The analysis and results for the proper
distances are given in appendix \ref{app:othermatchings}.

In light of these results, a different identification scheme is
required. Such a scheme has to be based on other diffeomorphism
invariant quantities, for example on curvature scalars. In
cosmological contexts, the Ricci scalar $R$ has been used
\cite{Hindmarsh:2012rc,Copeland:2013vva}.  However, the classical
Ricci scalar cannot be used, since it vanishes identically for vacuum
solutions of the Einstein field equations. Thus, in the following
analysis, we will base our scale identification on the Kretschmann
scalar $K=R_{\alpha\beta\gamma\delta}R^{\alpha\beta\gamma\delta}$, a
diffeomorphism invariant quantity of momentum dimension four. 
This motivates the scale identification 
\begin{equation}
D_\mathrm{K}(r)=\frac{1}{\chi\left(K-K_\infty\right)^{1/4}}\;,
\label{kretschmannident}
\end{equation}
with a constant $\chi$, chosen to be
$\chi=\left(\frac{1}{8}\right)^{1/4}$ in the following calculations,
and $K_\infty=K(r=\infty)=8/3\,\Lambda_0^2$ , using (\ref{kret}). We
subtract the Kretschmann scalar at $r\rightarrow\infty$, otherwise
$D(r)$ would approach a constant in the IR and therefore $G$ and
$\Lambda$ would fail to display the correct IR-limit $G_0$ and
$\Lambda_0$ respectively, cf.\ (\ref{analyticlambda}). For simplicity,
we base the matching on the classical Kretschmann scalar in the
equatorial plane ($\theta=\pi/2$). For both, 
Kerr-(A)dS and Schwarzschild-(A)dS we arrive at 
\begin{equation}
K=\frac{8}{3}\Lambda^2+
\frac{48M^2}{r^6}G^2\;.
\label{kret}
\end{equation}
The quantum improved version of the classical Kretschmann scalar
(\ref{kret}), referred to as $K_\mathrm{qu}$, provides a consistent
framework accounting for typical scales of the underlying (quantum)
geometry. Of course it would be desirable to use the true Kretschmann
scalar, computed directly from the quantum improved metric. This is
left for future work. On a technical level, the RG-improved version
turns (\ref{kretschmannident}) into a functional equation for
$D_\mathrm{K}(r)$. In order for this equation to have a
positive, real solution, $\chi$ must be constrained to
$\chi<\left(3/8\right)^{1/4}$, such that the expression under
the root in the UV-expression in \autoref{table_uvlimits_D} remains
positive. In appendix \ref{app:choiceofxi}, we discuss the impact of
$\chi$ on the results. Also, the quantum improved
version of classical Kretschmann scalar (\ref{kret}) approaches the
classical version for $r\rightarrow\infty$, but this does not hold for
$D_\mathrm{K}$, given by (\ref{kretschmannident}), because
$K_\mathrm{qu}\rightarrow K_\infty$ is faster than
$K_\mathrm{cl}\rightarrow K_\infty$. The
curvature near the singularity, the construction of the
Penrose diagrams, and the UV-limits for each proper distance are discussed
in appendix \ref{app:UVlimits}.

\begin{figure}[b]
	\includegraphics[width=0.49\textwidth]{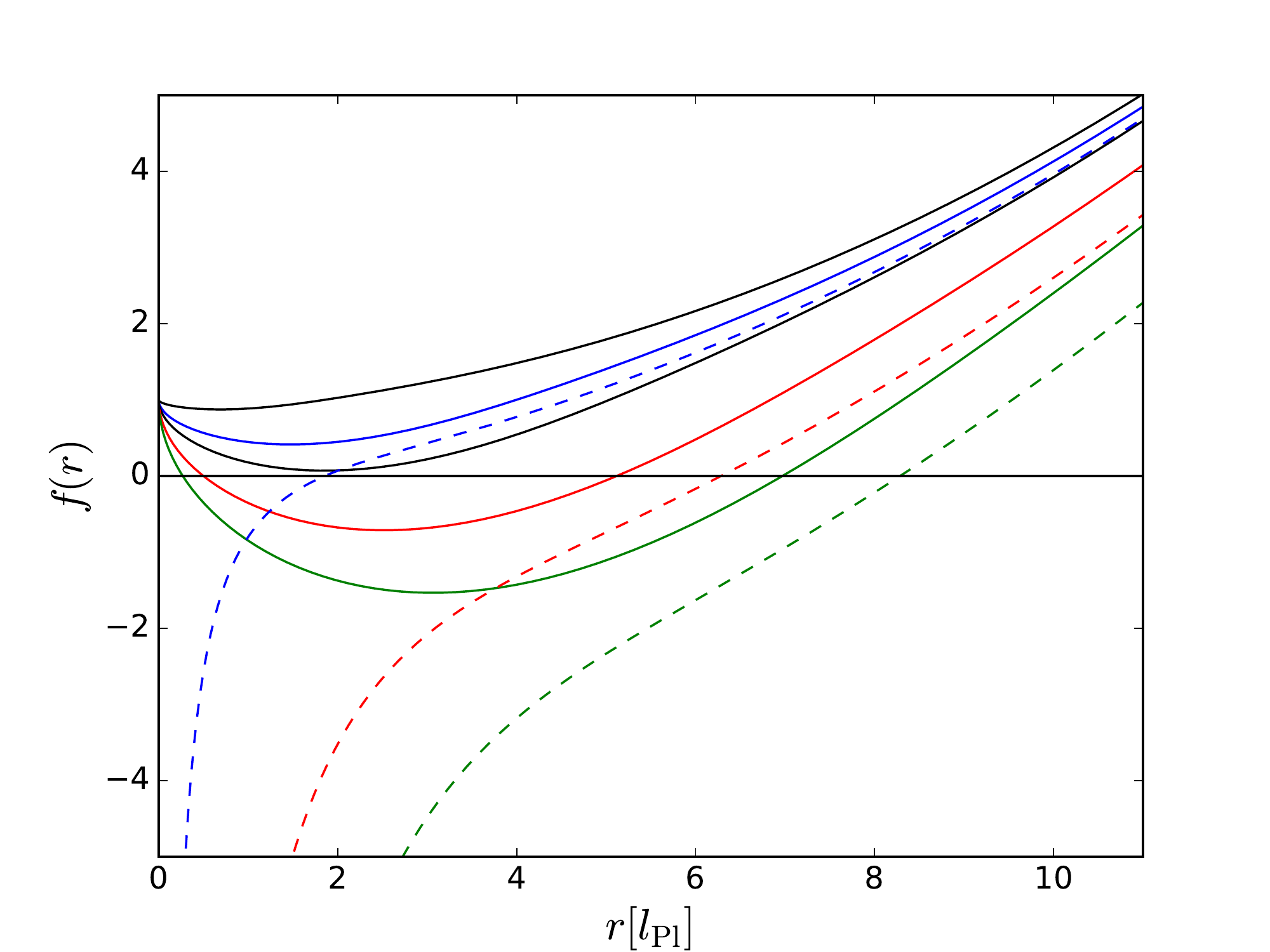}
	\caption{$f(r)$ from (\ref{schwads}) based on the Kretschmann
		scalar matching for increasing mass from top to
		bottom. Results based on the quantum improved Kretschmann
		scalar are given by solid curves, whereas results based on
		the classical Kretschmann scalar are dashed. The parameters are 
		$\Lambda_0=-0.1$ and $M=0.1,1,2,5,9M_\mathrm{Pl}$. Curves of
		the same mass have the same colour.}
	\label{fig:fsch_k}
\end{figure}

\section{Lapse Function and Number of Horizons}\label{sec:horizons}
With the running couplings $G$ and $\Lambda$ from the
previous section, physical properties of the quantum improved
space-times can be deduced. Central tools are the lapse functions
$f(r)$ and $\Delta(r)$, whose roots determine the location of horizons
in the space-time. These zeros are shown to be Killing horizons in
appendix \ref{app:killinghorizon}, implying that they can be assigned
a constant surface gravity, which turns out to be proportional to the
first derivative of the lapse function evaluated at the horizon. This
can be used to address thermodynamical processes such as the endpoint
of black hole evaporation via Hawking radiation. Another interesting question
is that of the similarity of the quantum improved geometry to the
classical geometry in general relativity, serving as a metric ansatz
for the quantum improvement. 

In this section, we will discuss the
lapse functions $f(r)$ and $\Delta(r)$ for the Kretschmann matching by
determining the number of horizons and comparing them with the lapse
functions of general relativity. We first start with asymptotically AdS
space-times, i.e. $\Lambda_0<0$, and comment on the results for
$\Lambda_0>0$ subsequently. The results for all other matchings can be
found in appendix \ref{app:othermatchings}.

\begin{figure}[h!]
	\includegraphics[width=0.49\textwidth]{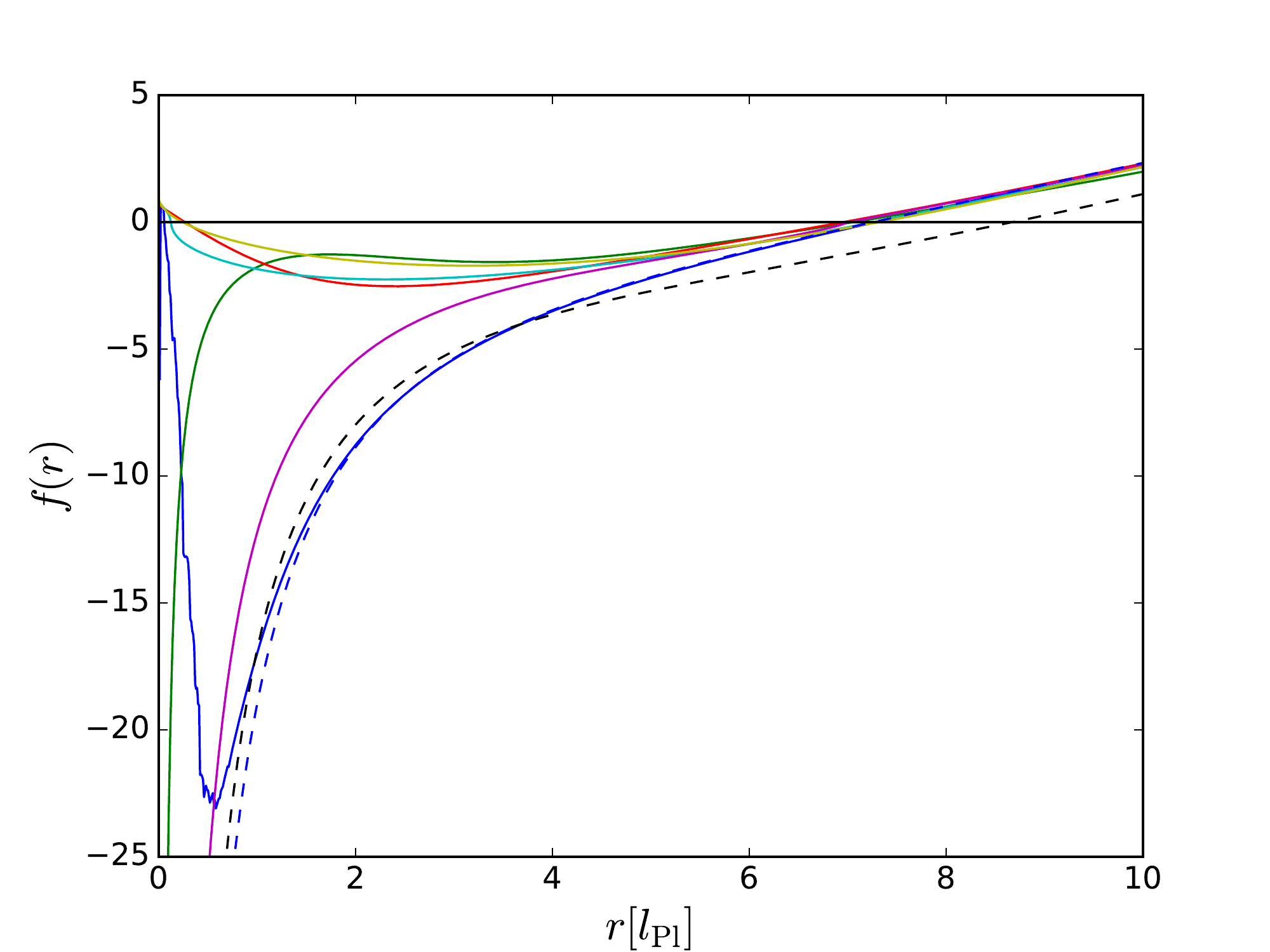}
	\caption{Comparison of $f(r)$ for all matchings with the
		classical result from general relativity for
		$M=10M_\mathrm{Pl}$ and $\Lambda_0=-0.1$. Matching based on
		the quantum geodesic in dark blue, classical geodesic in
		dark green, quantum radial path in light blue, classical
		radial path in purple, quantum Kretschmann scalar in light
		green, classical Kretschmann scalar in dashed black, linear
		matching in red and the result from general relativity in
		dashed dark blue. All matchings, apart from the classical
		Kretschmann setting, agree with the classical position of
		the outer black hole horizon.}
	\label{fig:fsch_classcomp}
\end{figure}

\subsection{Schwarzschild-AdS}

Classically, i.e. for constant $G$ \& $\Lambda_0<0$, the lapse
function $f(r)$ shows just one zero corresponding to the event horizon
of the black hole, whereas the quantum improved Schwarzschild geometry
shows up to two horizons, if a consistent matching is adopted,
\autoref{fig:fsch_k}. Starting at very large masses, well above the
Planck mass, we find two horizons, generated by a minimum of the lapse
function. Comparing with the classical lapse function in
\autoref{fig:fsch_classcomp} shows that the outer horizon of the
quantum improved space-time coincides with the event horizon of the
classical black hole. The larger the mass, the better the agreement
and the more the inner horizon moves towards zero. Hence, increasing
the mass makes the black holes more classical. Decreasing the mass
causes the minimum to shrink and the horizons to move towards each
other. There exists a critical mass $M_\mathrm{c}$ around two Planck
masses, $M_\mathrm{c}\approx 2 M_\mathrm{Pl}$, when the minimum is
also a zero of the lapse function. Then, both horizons merge and
$f(r)$ has a double root. We will see later, that this geometry is
similar to a classical, extreme Reissner-Nordstr\"om black hole in
AdS. For masses below the critical mass, the minimum is above zero and
no horizons are present.
\newpage
The results for matchings computed in space-times with running
couplings agree with the matchings based on space-times with constant
couplings on the position of the outer horizon, but differ
significantly for smaller radii. These differences emerge because in
the latter case, the matching is based on a classical geometry,
whereas we actually study a quantum geometry with running
couplings. Varying the amplitude for negative $\Lambda_0$ does not
affect the qualitative results, but changes the scale.

\subsection{Kerr-AdS}
\begin{figure}
	\includegraphics[width=0.5\textwidth]{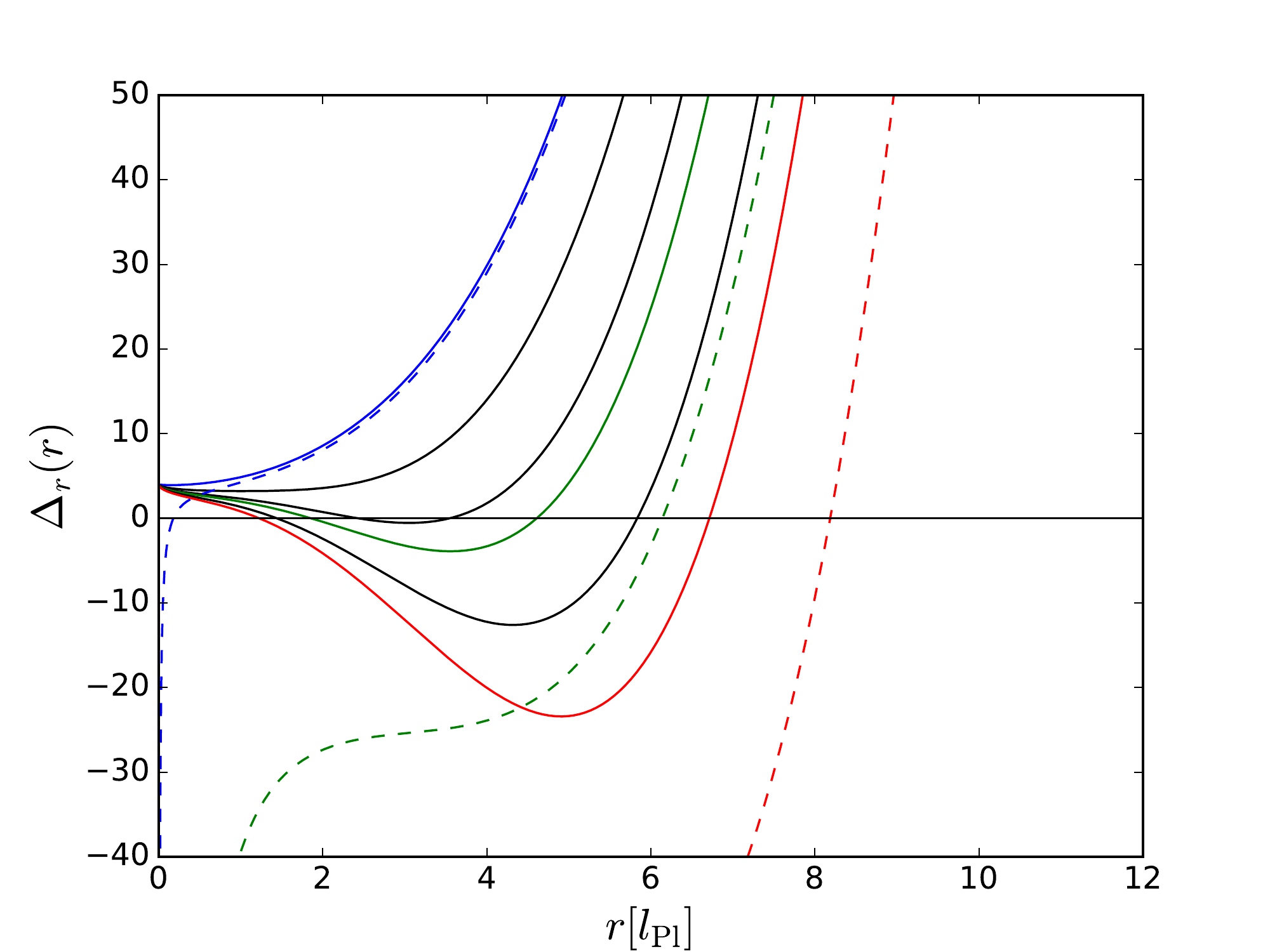}
	\caption{$\Delta_r(r)$ from (\ref{definitions}) based on the
          Kretschmann scalar matching for increasing mass from top to
          bottom. Results based on the quantum improved Kretschmann
          scalar are given by solid curves, whereas results based on
          the classical Kretschmann scalar are dashed. With parameters
          $\Lambda_0=-0.1$, $a=2$ and
          $M=0.1,2,4,5,7,9M_\mathrm{Pl}$. Curves of the same mass have
          the same colour.}
	\label{fig:delta_k}
\end{figure}

\begin{figure}
	\includegraphics[width=0.45\textwidth]{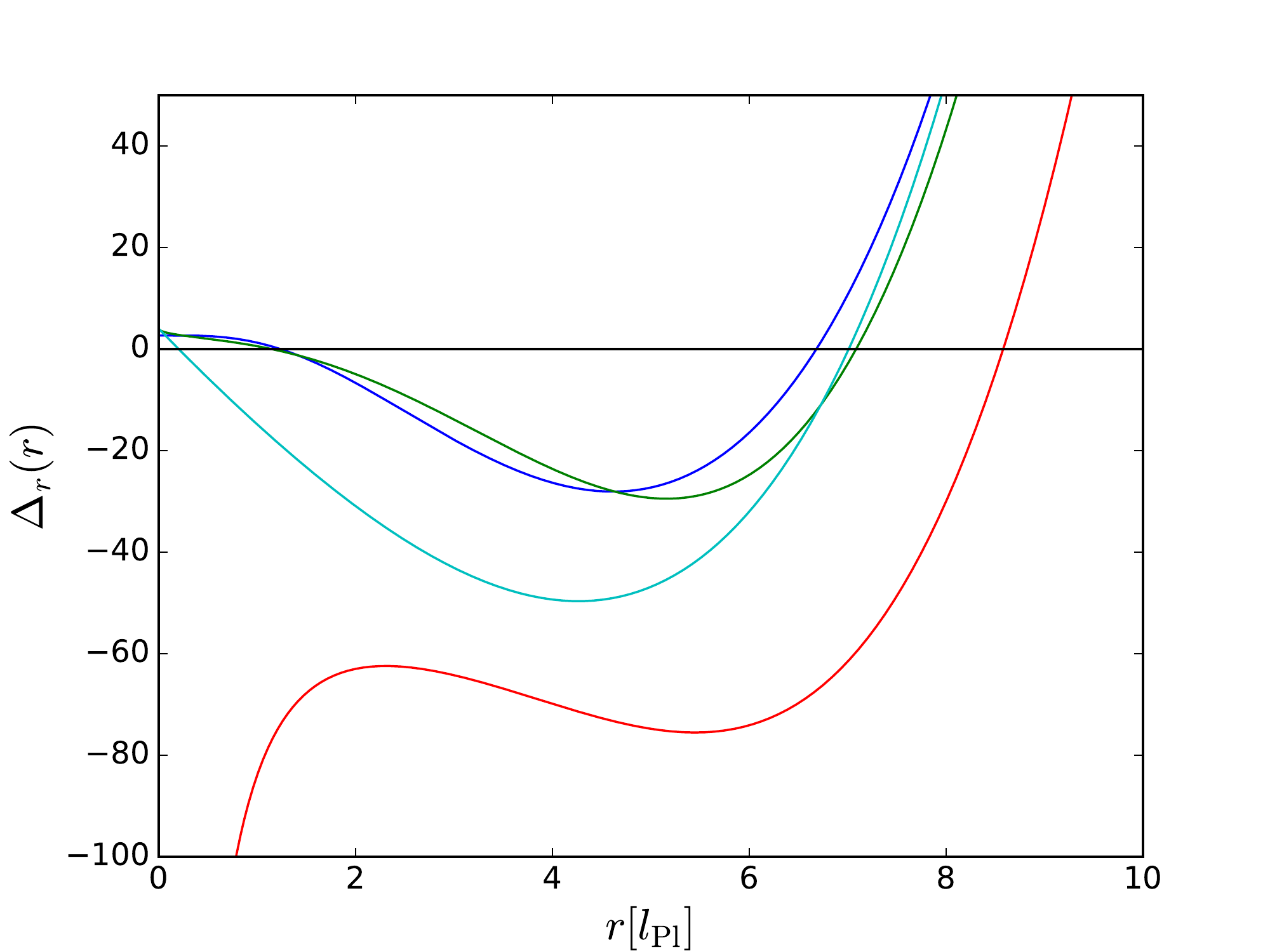}
	\caption{Comparison of $\Delta_r(r)$ for the linear matching in
		dark blue, the classical Kretschmann setting in red, the
		quantum Kretschmann setting in green and the classical
		result from general relativity in light blue, with
		$M=10M_\mathrm{Pl}$, $a=2$ and $\Lambda_0=-0.1$. Apart from
		the classical Kretschmann setting, all other matchings agree
		with the classical position of the outer horizon.}
	\label{fig:delta_classcomp}
\end{figure}

A classical, non-extremal Kerr-AdS space-time has two horizons: a
Cauchy horizon inside the black hole event horizon. In contrast to the
Schwarzschild case discussed above, the quantum improvement of this
space-time does not allow for more horizons than in the classical
geometry. Since the proper distances vanish identically in the
consistent scenarios, we show only the results for the Kretschmann
matching in \autoref{fig:delta_k} and the dependence on the rotation
parameter for fixed mass in \autoref{fig:delta_a_k}. The results for
the linear matching can be found in appendix
\ref{app:othermatchings}. In general, the consistent quantum improved
version displays the same behaviour as the classical
solution. However, the inner horizon in the quantum improved
space-time is located at larger radii than the classical Cauchy
horizon, \autoref{fig:delta_classcomp}.

\begin{figure}
	\includegraphics[width=0.5\textwidth]{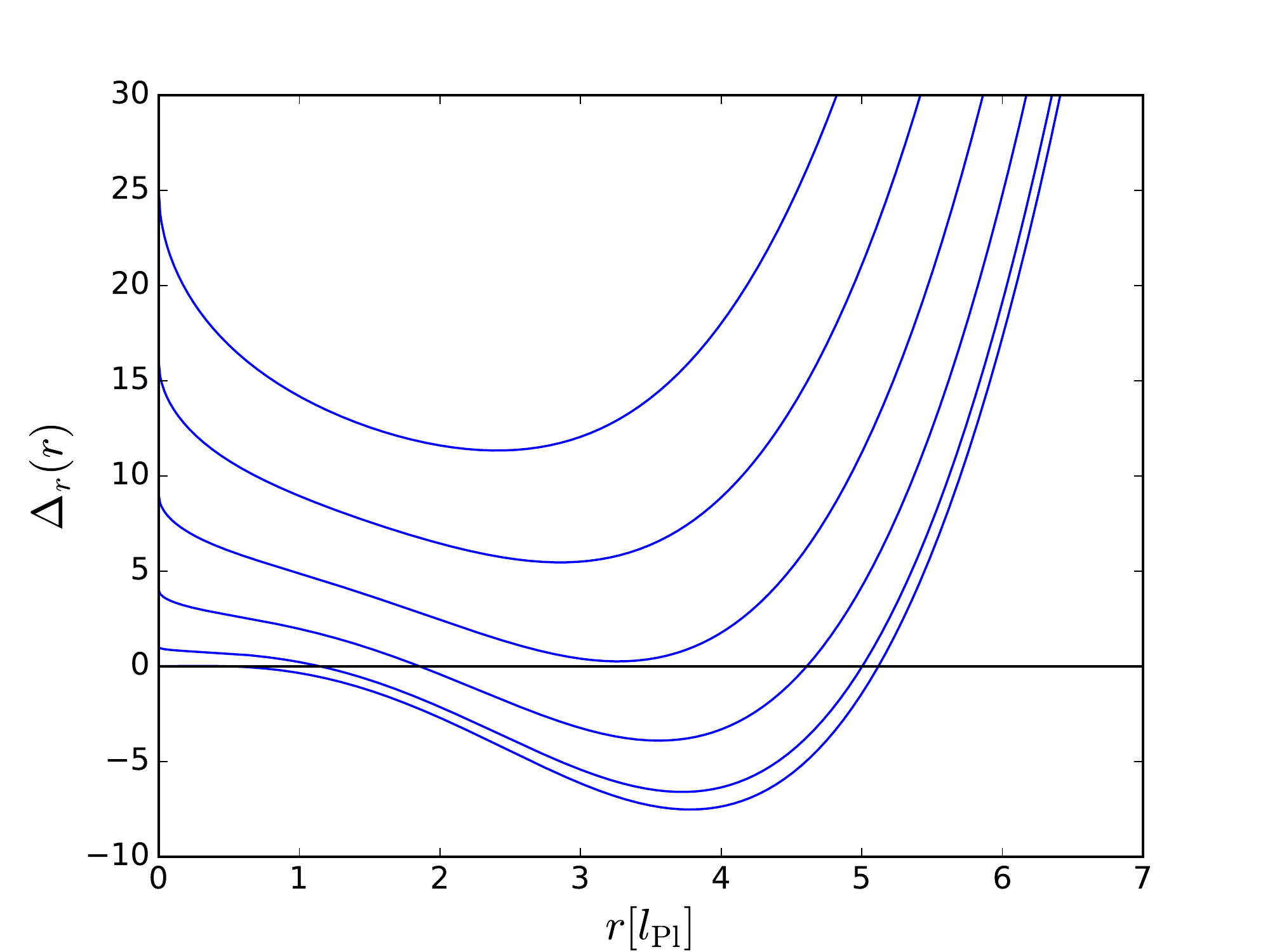}
	\caption{$\Delta_r(r)$ based on the quantum Kretschmann scalar
		matching for fixed mass $M=5M_\mathrm{Pl}$ and
		$\Lambda_0=-0.1$, but increasing $a=0,1,2,3,4,5$ from bottom
		to top.}
	\label{fig:delta_a_k}
\end{figure}

\subsection{Asymptotically de Sitter spaces}
\begin{figure}[h]
	\includegraphics[width=0.45\textwidth]{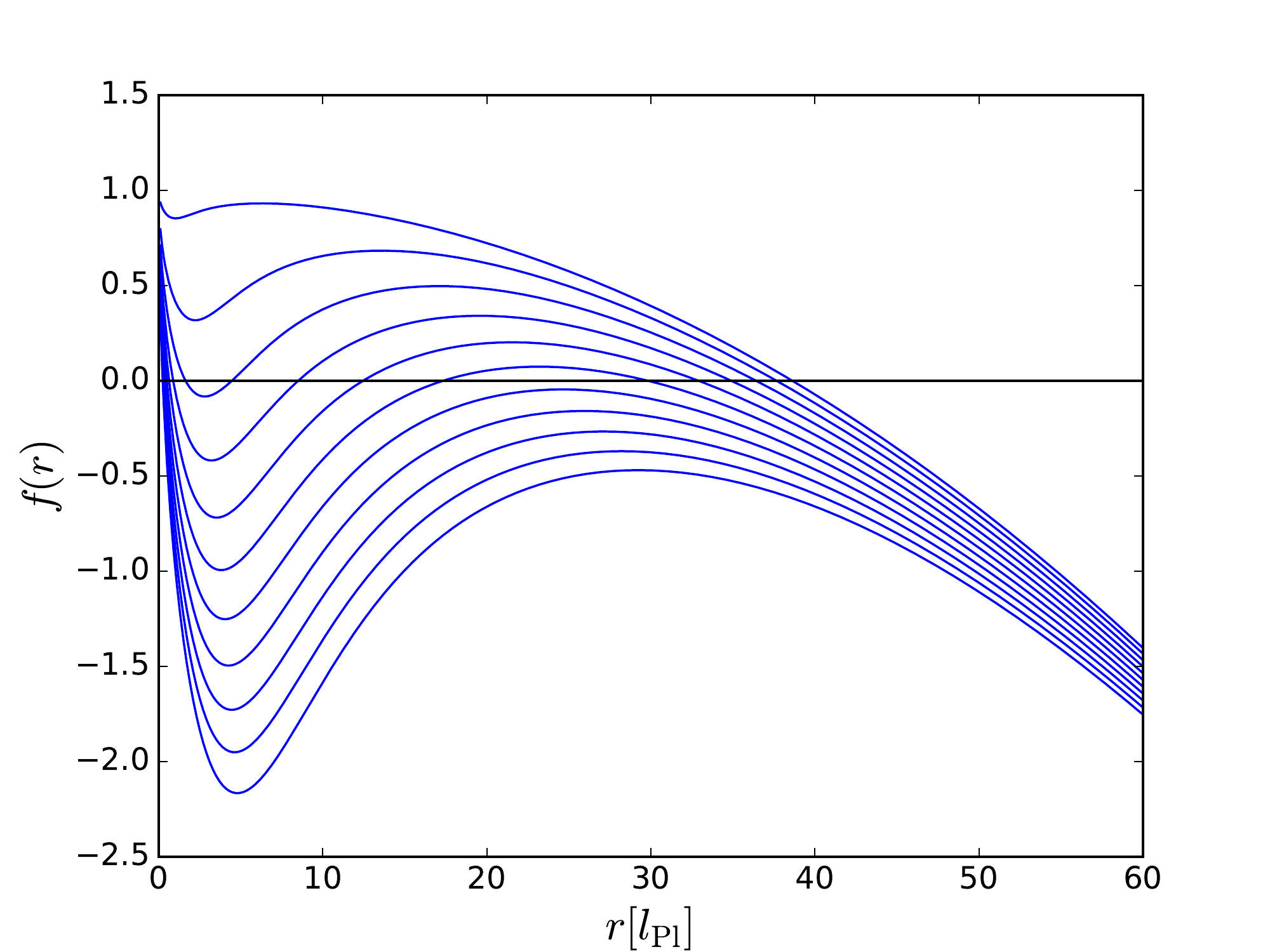}
	\caption{$f(r)$ for asymptotic dS with
          $\Lambda_0=0.001$ for increasing mass
          $M=0.1,1,2,3,4,5,6,7,8,9,10M_\mathrm{Pl}$ from top to
          bottom.}
	\label{fig:dS_fsch_kqu}
\end{figure}

\begin{figure}[h]
	\includegraphics[width=0.45\textwidth]{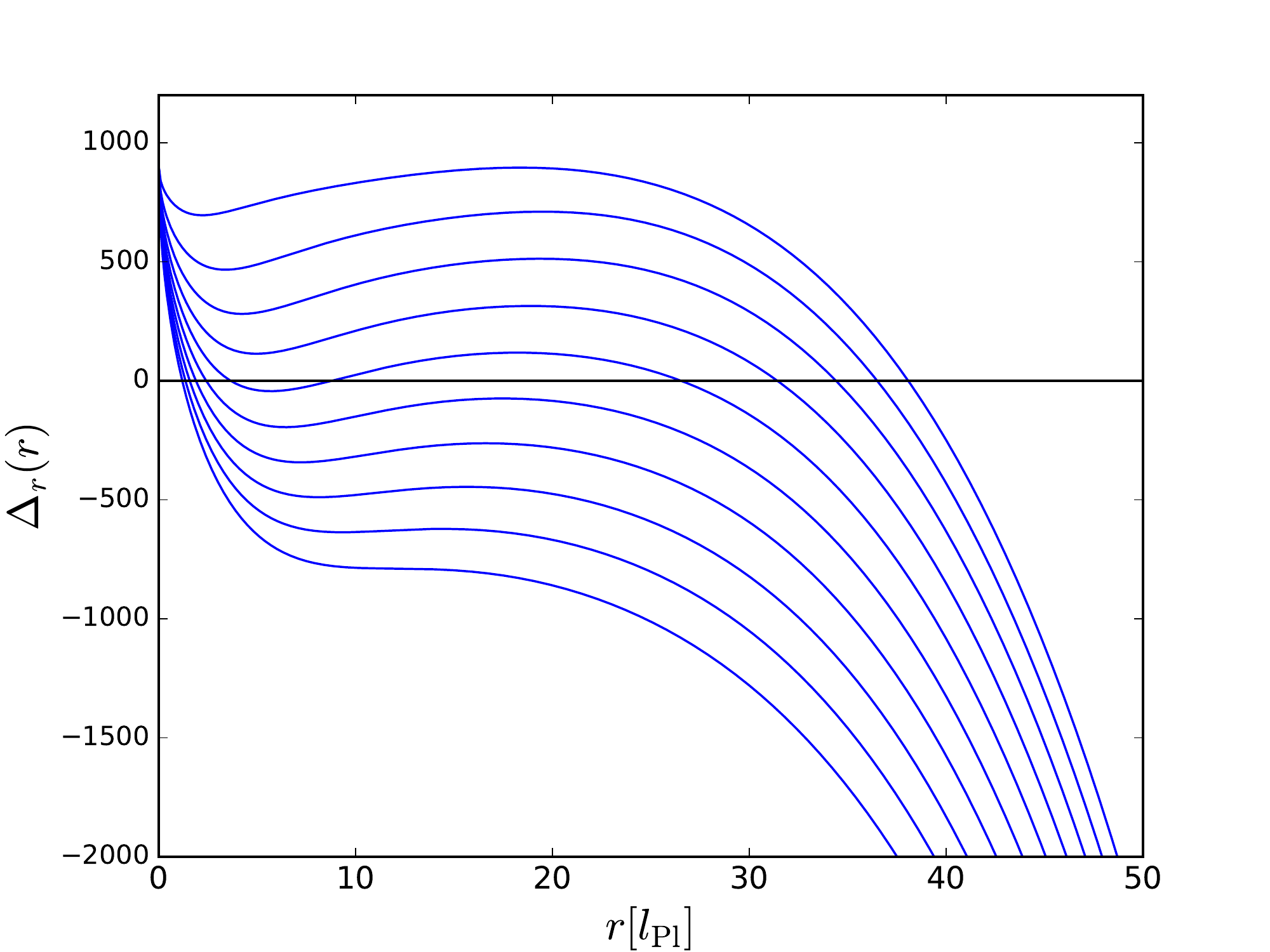}
	\caption{$\Delta_r(r)$ for asymptotic dS
		with $\Lambda_0=0.001$ and $a=30$ for increasing mass
		$M=1,3,5,7,9,11,13,15,17,19M_\mathrm{Pl}$ from top to
		bottom.}
	\label{fig:dS_delta_kqu}
\end{figure}

\begin{figure}[h]
	\includegraphics[width=0.45\textwidth]{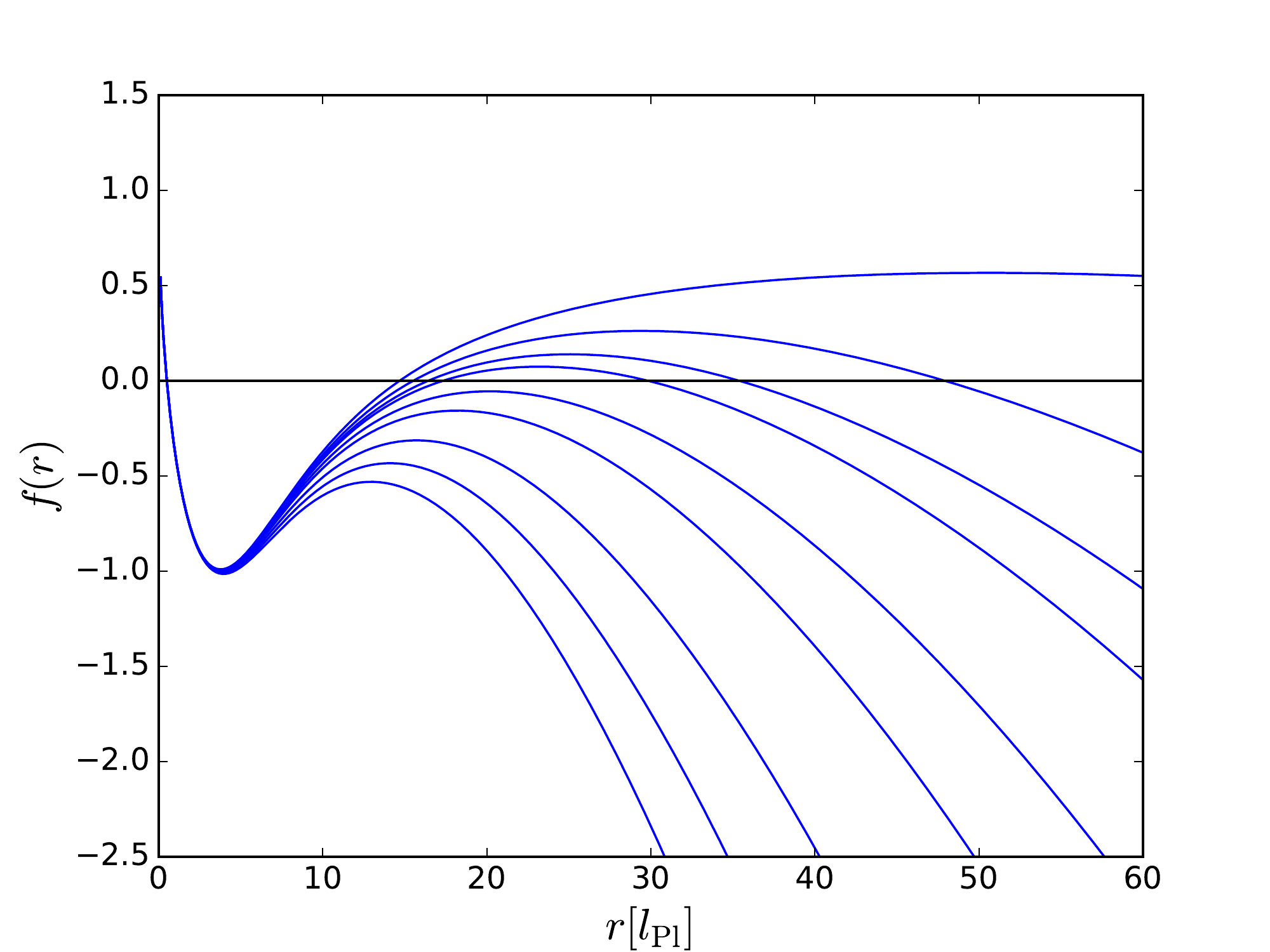}
	\caption{$f(r)$ for asymptotic dS for
          increasing
          $\Lambda_0=0.0001,0.0005,0.0008,0.001,0.0015,0.002,0.003,0.004,0.005$
          from top to bottom and fixed mass $M=5M_\mathrm{Pl}$.}
	\label{fig:dS_fsch_varLam}
\end{figure}

\begin{figure}[b]
	\includegraphics[width=0.45\textwidth]{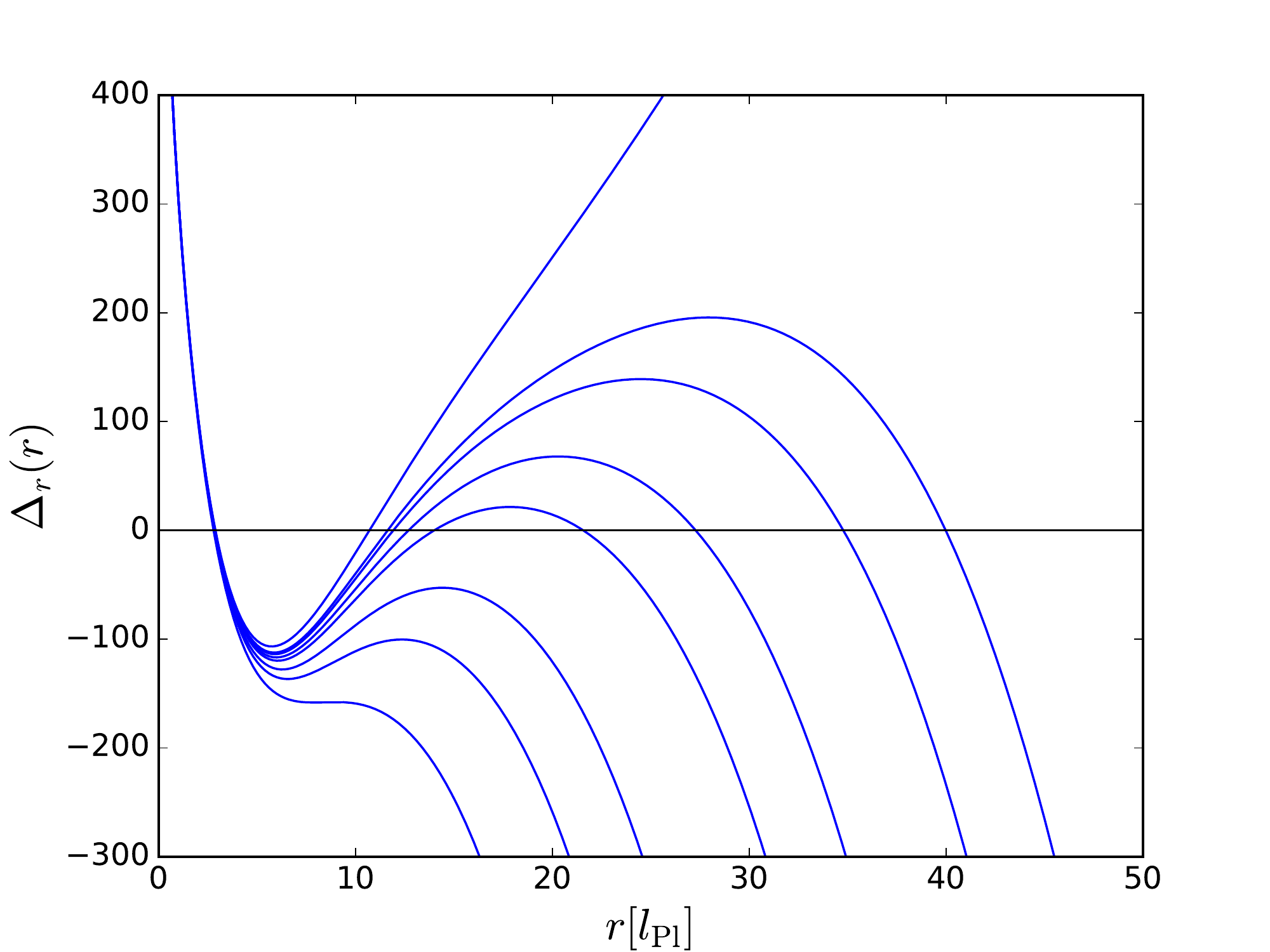}
	\caption{$\Delta_r(r)$ for asymptotic dS for
          increasing
          $\Lambda_0=0.0001,0.0005,0.0006,0.0007,0.0009,0.0015,0.002,0.003$
          from top to bottom. Fixed mass $M=5M_\mathrm{Pl}$ and
          $a=30$.}
	\label{fig:dS_delta_varLam}
\end{figure}

If we take the space-time to be asymptotically de Sitter, we find the
possibility to get up to three horizons. The additional horizon is
generated by the positive cosmological constant in the IR and appears
in the classical regime at large radii. The typical shapes of $f(r)$
and $\Delta_r(r)$ are displayed in \autoref{fig:dS_fsch_kqu} \&
\autoref{fig:dS_delta_kqu} for the Kretschmann matching, the dependence on
the amplitude of $\Lambda_0$ is shown in
\autoref{fig:dS_fsch_varLam} \& \autoref{fig:dS_delta_varLam}. Varying
$m$ controls the position of the two inner 
horizons via the formation
of a minimum, whereas $\Lambda_0$ governs the location of the outer
horizon. Thereby, the interplay of the amplitudes of $m$ and
$\Lambda_0$ dictates the number of horizons. Although we cannot
provide an analytical condition involving $m$ and $\Lambda_0$ for the
space-time exhibiting three horizons, it is suggestive to see it 
as the generalised version of the condition for a classical
Kerr-dS space-time to have three horizons. This also implies that both quantum improved
space-times have two distinct extremal cases: both inner horizons
merge at a mass $m=M_*$ yielding an extremal black hole inside the
cosmological horizon. Or both outer horizons merge at $m=M^*$, forming
the largest Schwarzschild/Kerr-dS black hole possible, analogous to the
Nariai space-time.
\enlargethispage{2\baselineskip}

\clearpage
\newpage

\section{Global Structure, Penrose Diagrams and Particle Trajectories}\label{sec:penrose}
In contrast to the classical Schwarzschild-(A)dS and Kerr-(A)dS
geometries of general relativity, the quantum improved counterparts
can exhibit a different number of horizons and hence may show a
different global structure, depicted in terms of Penrose diagrams.  It
turns out that both geometries, i.e. one based on the Schwarzschild
and the other on the Kerr metric, have the same Penrose diagram. The
resulting diagram is equivalent to the classical Reissner-Nordstr\"om
or Kerr geometry. Hence, the quantum improvements of the metric lead
to a unified global structure for quantum improved black hole
space-times based on solutions of the Einstein field equations. Yet,
as it is shown in section \autoref{sec:trajectories} below, particles move
differently in each geometry.

We start by determining whether the singularity is time-like,
space-like or null. To that end we compute the norm of the normal
vector of a hypersurface of constant $r$ in the limit
$r\rightarrow 0$. The norm turns out to be the $rr$-element of the
inverse metric $g^{rr}$, yielding
\begin{equation}
  g^{rr}_\mathrm{Sch}\overset{r\rightarrow 0}{=}1 \qquad \&\qquad 
  g^{rr}_\mathrm{Kerr}\overset{r\rightarrow 0}{\rightarrow}\frac{1}{\cos^2 \theta} \;.
\end{equation}
Hence, the singularity is time-like in both cases, irrespective of
whether the space-time is asymptotically AdS or dS. As it is shown in
appendix \ref{app:killinghorizon}, zeros of $f$ and $\Delta_r$
correspond to Killing horizons. The succession of sign changes of the
lapse function dictates how the hypersurfaces of constant $r$ change
from time-like over null to space-like.

\subsection{Asymptotically anti-de Sitter space-times}
\begin{figure*}[th]
	\includegraphics[width=0.28\textwidth]{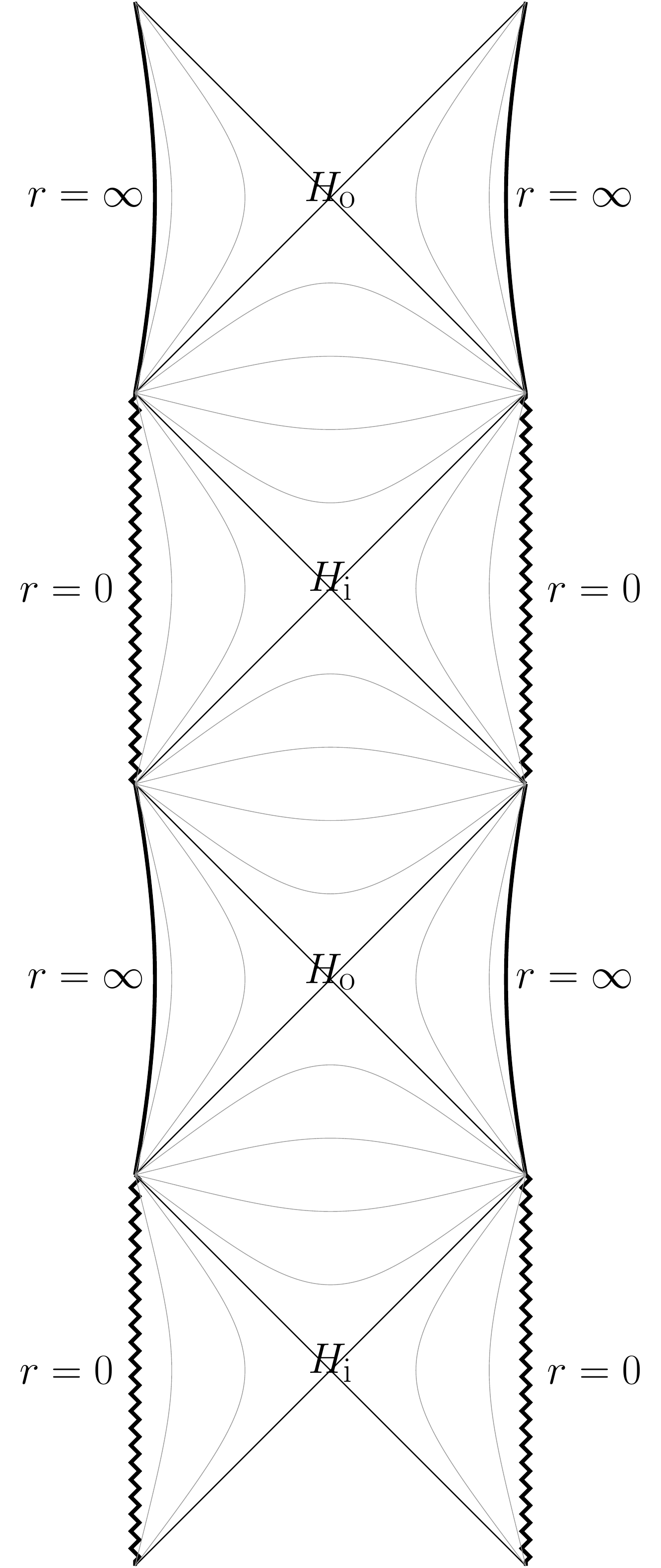}\hfill
	\includegraphics[width=0.28\textwidth]{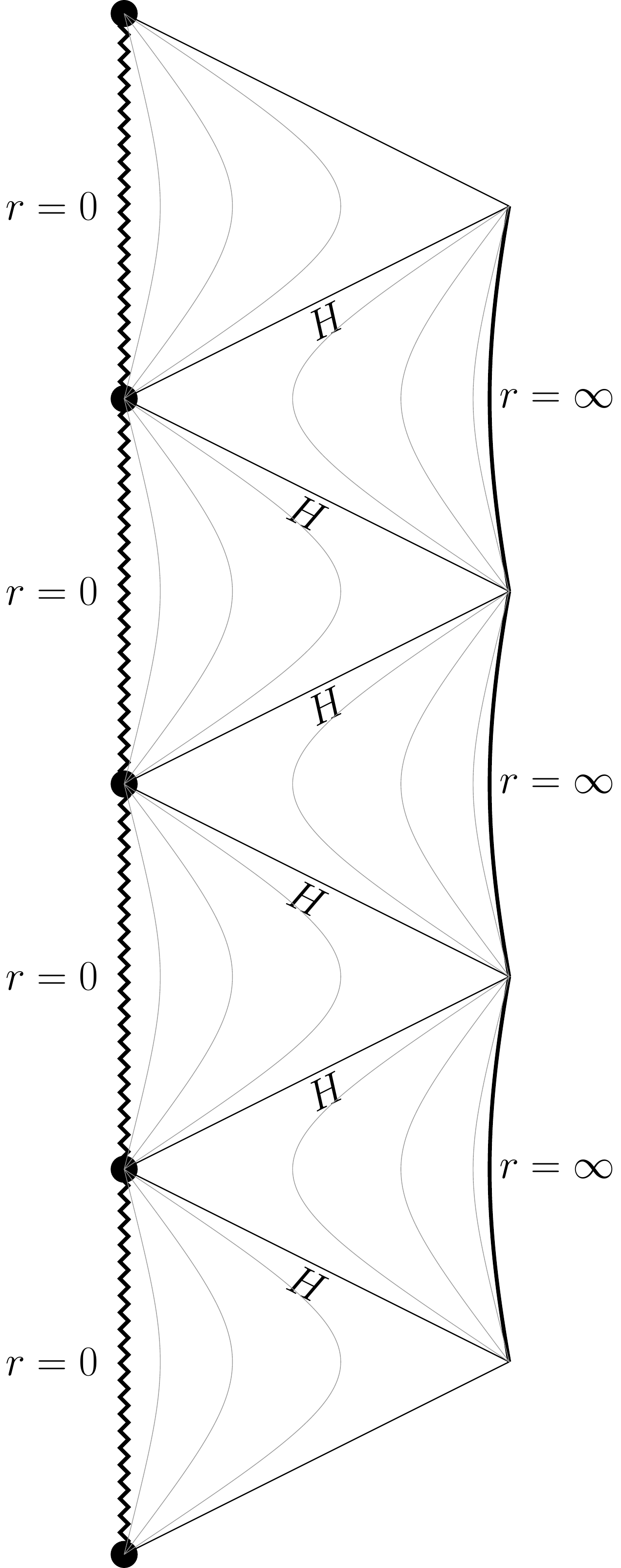}\hfill
	\includegraphics[width=0.28\textwidth]{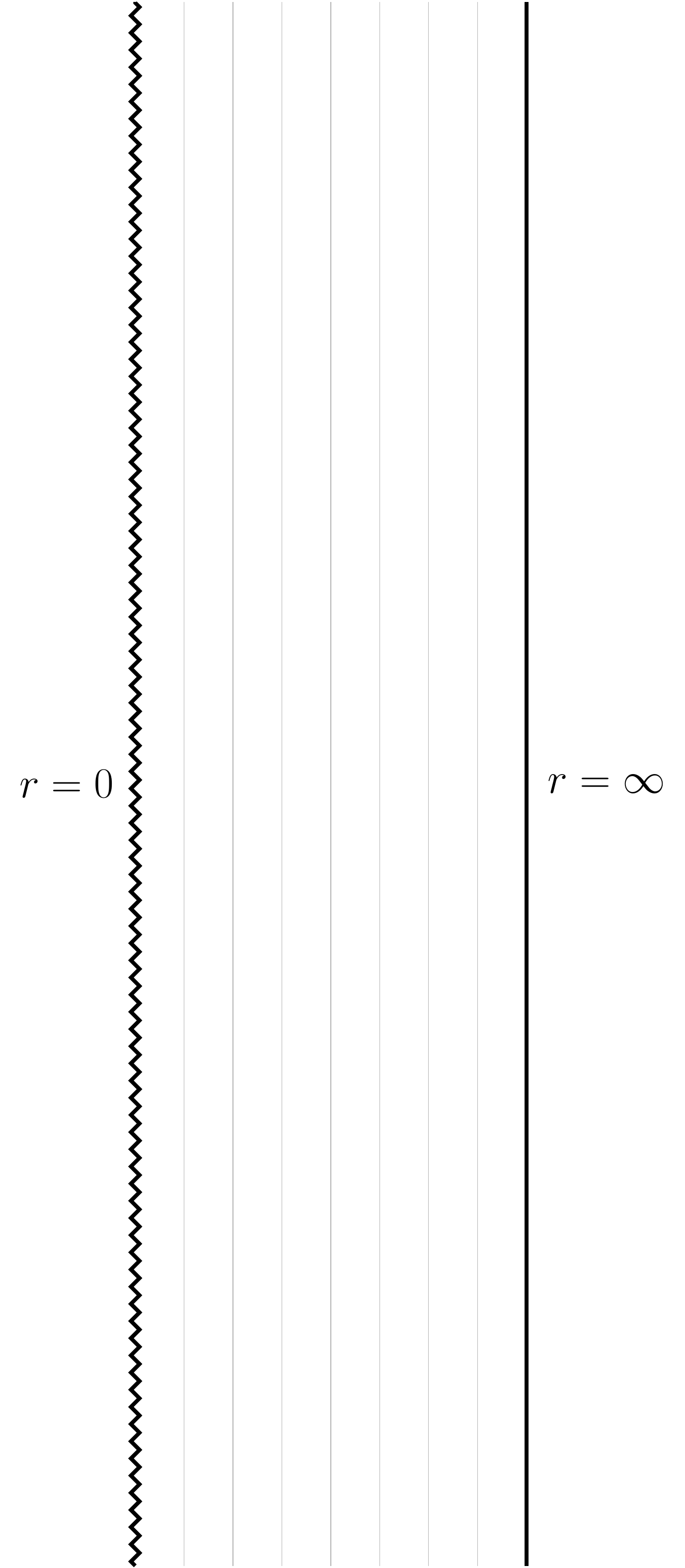}
	\caption{Penrose diagrams for quantum improved Schwarzschild- and Kerr-AdS space-times. hypersurfaces $r=$const. are drawn in grey, each diagram can be further extended in vertical direction. To the left the Penrose diagram for the non-extremal black hole with outer horizon $H_\mathrm{o}$ and inner horizon $H_\mathrm{i}$, the timelike singularity ($r=0$) and conformal infinity ($r=\infty$). In the middle the diagram for the extremal geometry with just one horizon $H$. The black dots are not part of the singularity. To the right, the diagram for AdS with a naked singularity at $r=0$.}
	\label{fig:penrose_ads}
\end{figure*}

The lapse function of Schwarzschild-AdS and the Kerr-AdS space-time
share the same qualitative features, resulting in the same Penrose
diagram. The formal construction of the maximally extended space-time
works the same as for the classical Kerr space-time, for instance see
\cite{Carter:1968rr,Gibbons:1977mu}, but now with an asymptotic
AdS-patch. For a mass larger than the critical mass $M_\mathrm{c}$,
the lapse function has two distinct roots, so the space-time exhibits
two horizons, \autoref{fig:penrose_ads}. When $m=M_\mathrm{c}$, both
roots coincide and we find an extremal black hole with just one
horizon.  For even lower masses, that is $m<M_\mathrm{c}$, no horizon
is present, but the singularity still exists,
cf.~\autoref{sec:singularity}, leaving a space-time with a naked
singularity. Later, via a heuristic argument, we will argue that this
unphysical space-time cannot be formed by gravitational collapse.

\subsection{Asymptotically de Sitter space-times}
\begin{figure*}
	\includegraphics[scale=0.35]{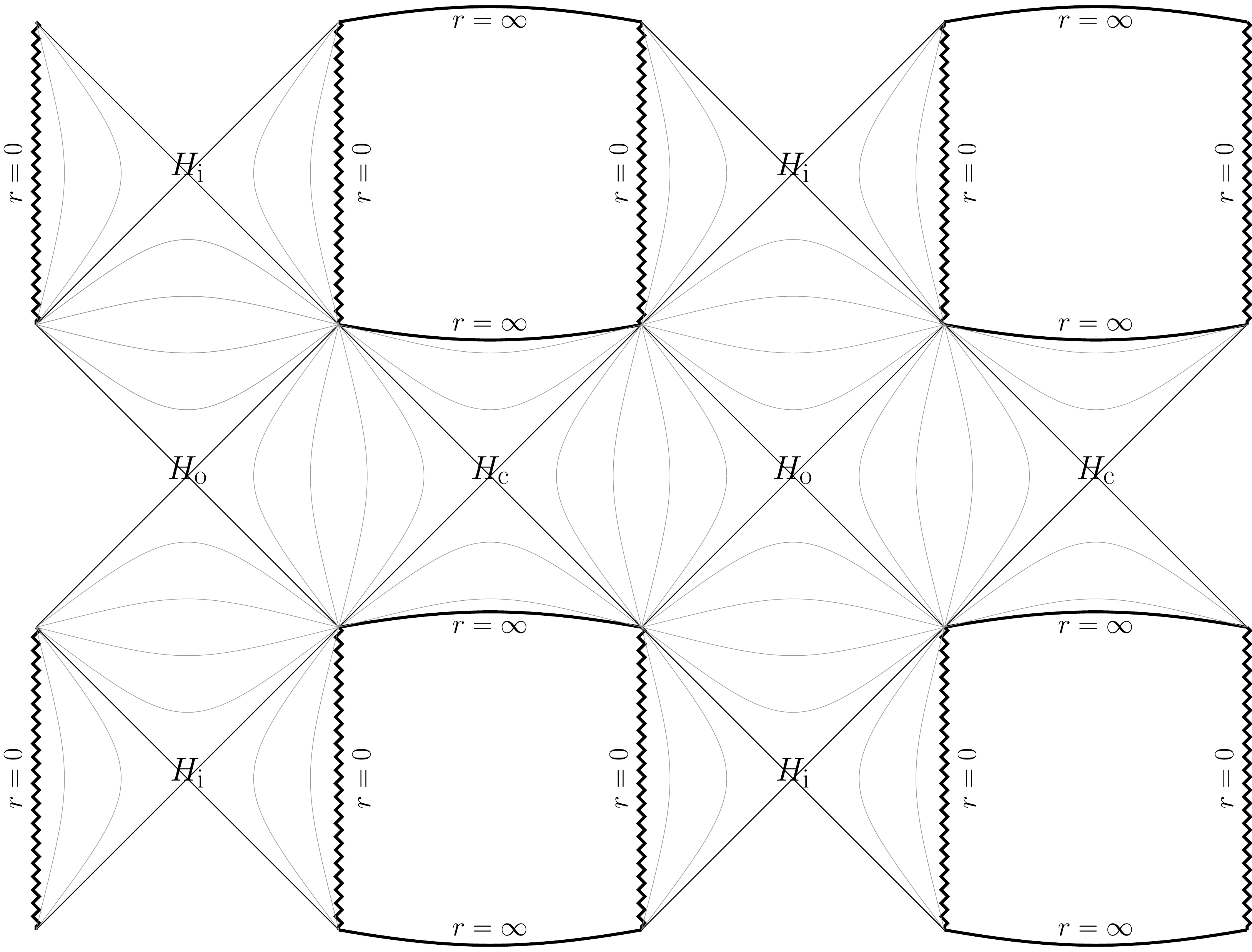}
	\caption{Penrose diagram for quantum improved Schwarzschild- and Kerr-dS geometry with the three horizons of a non-extremal black hole configuration. Starting at the timelike singularity at $r=0$, we first cross the inner horizon $H_\mathrm{i}$ and then the outer horizon $H_\mathrm{o}$ before crossing the cosmological horizon $H_\mathrm{c}$ and reaching conformal infinity $r=\infty$. This diagram can be further extended into all directions. Again, $r=$const. hypersurfaces are depicted by grey curves.}
	\label{fig:penrose_ds_threehor}
\end{figure*}

\begin{figure}
	\includegraphics[width=0.44\textwidth]{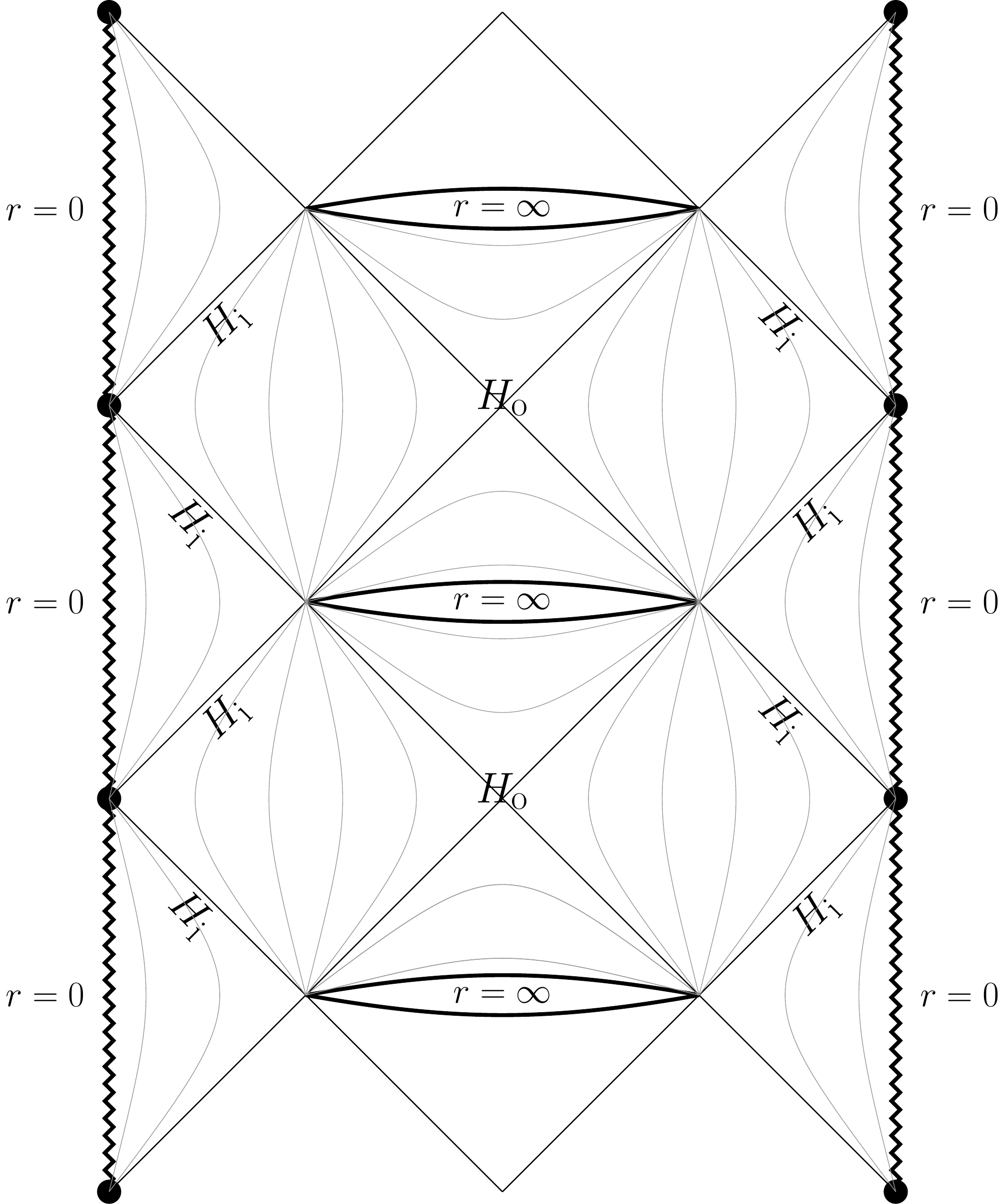}
	\caption{Penrose diagram for quantum improved Schwarzschild- and Kerr-dS geometry with the two horizons of an extremal black hole configuration. Starting at the curvature singularity at $r=0$, we first cross the inner horizon $H_\mathrm{i}$ and then the outer one $H_\mathrm{o}$, before arriving at conformal infinity $r=\infty$. This diagram can be further extended to the top and bottom as well. The black dots are not part of the singularity. The displayed pattern of the $r=$const. hypersurfaces is the one for $m=M^*$. For $m=M_*$, the hypersurfaces between the horizons become spacelike.}
	\label{fig:penrose_ds_twohor}
\end{figure}

\begin{figure}
	\includegraphics[scale=0.47]{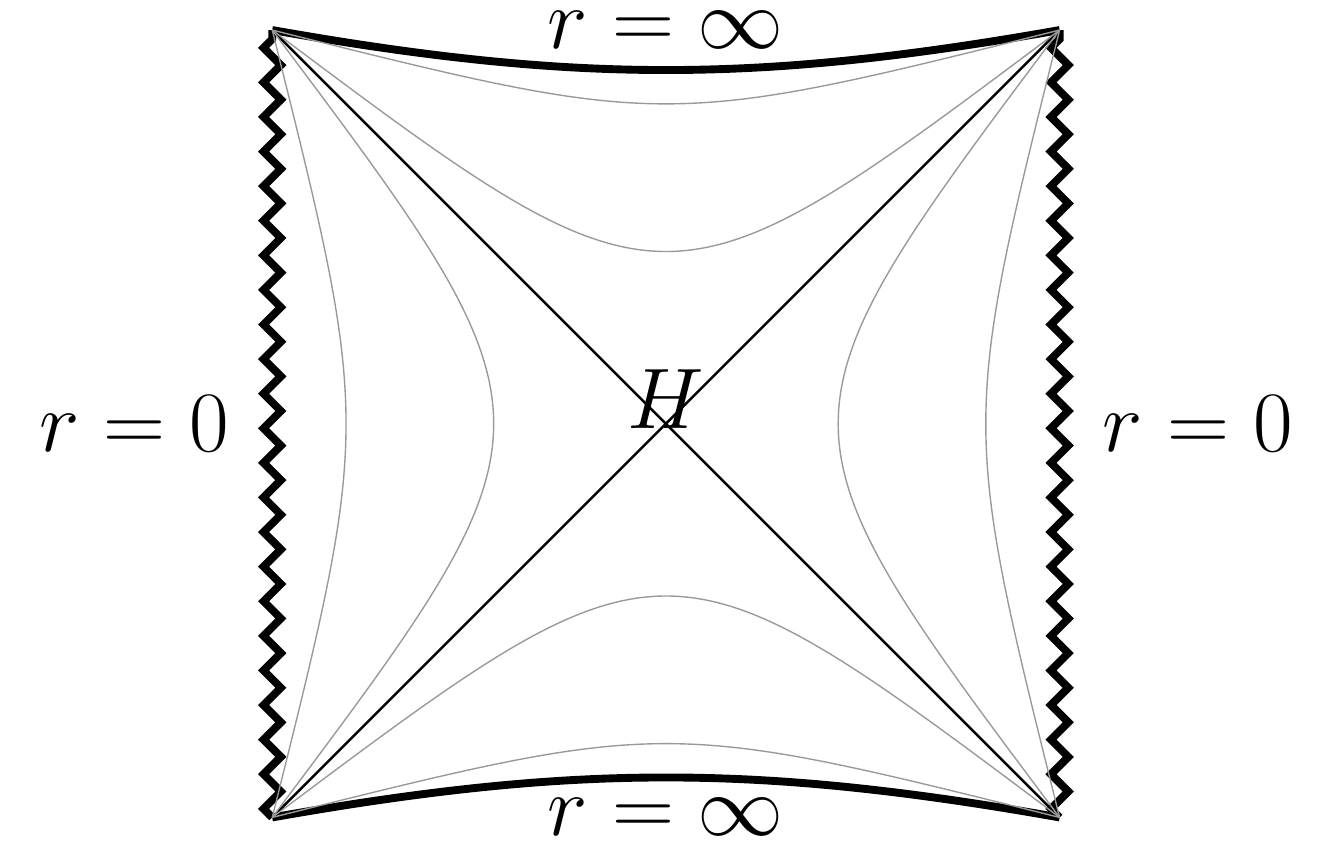}
	\caption{Penrose diagram for the quantum improved Schwarzschild- and Kerr-AdS configuration showing only one horizon $H$, always shielding the singularity at $r=0$ from an observer near conformal infinity $r=\infty$.}
	\label{fig:penrose_ds_onehor}
\end{figure}

The results for the Schwarzschild- and Kerr-dS geometries agree with
each other.  The space-time exhibits two distinguished masses,
$M_*<M^*$, at which two of the possible three horizons merge.
Starting with $M_*<m<M^*$, the space-time has three distinct horizons,
two of them are associated with the black hole and one with the
positive cosmological constant on large scales,
\autoref{fig:penrose_ds_threehor}. This case is equivalent to the
classical Kerr-dS geometry. For $m=M^*$, the outer black hole horizon
and the cosmological horizon merge. This leaves an extremal space-time
containing a maximally sized black hole,
\autoref{fig:penrose_ds_twohor}, similar to the Nariai space-time. For
even larger masses, there is just one horizon left,
\autoref{fig:penrose_ds_onehor}. On the other end, the de Sitter
space-time contains an extremal black hole if $m=M_*$.  For $m<M_*$,
we have a de Sitter geometry containing singularity, which is naked
for observers within the cosmological horizon. The construction of the
maximally extended space-time is analogous to the one for the
classical Kerr-dS case, described for instance in
\cite{Gibbons:1977mu}.


\subsection{Particle Trajectories}
\label{sec:trajectories}
In order to investigate whether particles propagate differently in the
quantum space-times as compared to general relativity, we study their
trajectories. Although most new effects in quantum improved
space-times happen around the Planck scale, there are possibly deviations
from classical trajectories already on length scales well above. Our set-up
in the following is a test mass with zero angular momentum $L$ along
its (timelike) geodesic in a non-extremal geometry, neglecting all
backreactions. Furthermore, we are allowed to restrict the motion to
the equatorial plane, see \cite{Hackmann:2010zz} for more details. In order
to classify orbits into categories, for instance orbits terminating at
the central curvature singularity or bound ones, it suffices to study
only the change of the radial coordinate.

\subsubsection{Schwarzschild}
In the quantum improved Schwarzschild geometry, the equation for the
radial motion of a test mass, starting with zero angular momentum
$L$ at some distance $r$ with energy $E$, reads according to
(\ref{geodsch})
\begin{align}
\dot{r}^2=E^2-f(r)\;,
\end{align}
where $\dot{r}$ denotes the change of the radial coordinate along the
geodesic parametrised by the eigentime. This equation is only
dependent on $r$ and can be thought of as an energy equation per unit
mass for the total energy $E$ of the test particle in an effective,
one-dimensional potential $f(r)$. As was already found in \cite{Bonanno:2000ep}
for the asymptotically flat case, possible trajectories are the same
as in the classical Reissner-Nordstr\"om scenario, thereby differing
significantly from a classical Schwarzschild set-up. The only
difference to the asymptotical flat case arises at large scales, where
the effective potential $f(r)\rightarrow\pm\infty$, depending on
whether the space-time is asymptotically de Sitter or anti-de
Sitter. Recalling the shape of $f(r)$, e.g.\ \autoref{fig:fsch_k}, we
note that the effective potential is repulsive close to the
singularity. In an asymptotically AdS geometry and for a test mass
with energy $E$, the following options are possible, all being bound
orbits in radial direction:
\begin{enumerate}
\item If $E$ equals the minimum of the lapse function
  $f_\mathrm{min}$, then the particle is on a circular, stable orbit
  in the region between the horizons. The radius is determined by the
  distance where the repulsive singularity balances the repulsive
  negative asymptotical cosmological constant.
\item For $f_\mathrm{min}<E<0$, the particle is on a bound orbit,
  remaining in the region between both horizons.
	\item If $0<E<1$, the orbit will again be bound, but now the
          particle periodically crosses horizons. For instance, first
          starting in the region outside of the outer horizon, the
          trajectory will first cross the outer horizon, then the
          inner one. Because it cannot overcome the repulsive barrier
          of the singularity, it is bounced back and the radius is
          increasing again. By crossing another horizon, it will end
          up in an identical patch of the extended space-time. This
          motion continues indefinitely and the particle will travel
          through infinitely many universes. We will comment on the
          physicality of this scenario at the end of this section.
	\item If $E>1$, the energy of the particle can overcome the
          potential barrier and manages to approach the singularity at
          $r=0$ with non-zero kinetic energy. But in contrast to the
          classical Schwarzschild-AdS scenario, the particle again
          follows a path through infinitely many identical universes,
          reaching the singularity in each of them.
\end{enumerate}
For the case of a non-extremal black hole with asymptotic de Sitter
patch, we note that the maximum $f_\mathrm{max}$ is always smaller
than one. Therefore, we find scenarios one and two from above, but
also some differences:
\newpage
\begin{enumerate}
	\setcounter{enumi}{4}
      \item The case $0<E<f_\mathrm{max}$ admits a bound orbit, equivalent to scenario
        three with the outer turning point of the particle being located between
        the cosmological and the outer black hole horizon, as well as an unbound one beyond the cosmological horizon.
      \item For $E=f_\mathrm{max}$, the particle is at rest at the
        distance, where the attracting force of the black hole
        balances the attraction generated by the positive cosmological
        constant on large scales. This is an unstable equilibrium,
        since small perturbations cause the particle either to move
        inwards in a similar way to five, or to escape to infinity.
      \item In contrast to all above cases, the orbit is unbound in
        radial direction for $E>f_\mathrm{max}$, and the particle can
        escape to infinity. Depending on whether or not $E\gtrless 1$,
        it can reach the singularity at $r=0$.
\end{enumerate}

\subsubsection{Kerr}
The equation for the change of the radial coordinate along the
geodesic of a test particle with energy $E$ and zero angular momentum
$L$ in the equatorial plane of the Kerr geometry reads (cf. (\ref{radial_equ_kerr})), 
\begin{align}
\dot r^2 = R(r):= \frac{E^2\,\Xi^2\left[ (r^2+a^2)^2-a^2\Delta_r 
	\right]-r^2\Delta_r}{r^4}\;,
\label{Kerr_orbit}
\end{align}
where we introduced the function $R(r)$ for convenience. For a fixed
geometry $(G_0,\Lambda_0,M,a)$, the energy $E$ of the particle
determines the allowed orbits. In the following, we continue closely
along the more detailed analysis of the classical Kerr-(A)dS geometry
carried out in \cite{Hackmann:2010zz}. Since the above equation is quadratic
in $\dot r$, geodesics always have to satisfy $R(r)\ge 0$. A simple
root of $R(r)$ corresponds to a turning point, where the particle
comes to rest. A circular orbit of constant $r=r_0$
requires both $\dot r$ and $\ddot r$ to vanish at $r_0$, translating
via equation (\ref{Kerr_orbit}) into the condition of $R(r)$ having an
extremum as well as a zero at $r_0$. Depending on whether this
extremum is a maximum or minimum, the circular orbit will be stable or
unstable. Hence, $R(r)$ having at least a double zero at $r_0$ is a sufficient
condition for a circular orbit.

\begin{figure}
	\includegraphics[width=0.49\textwidth]{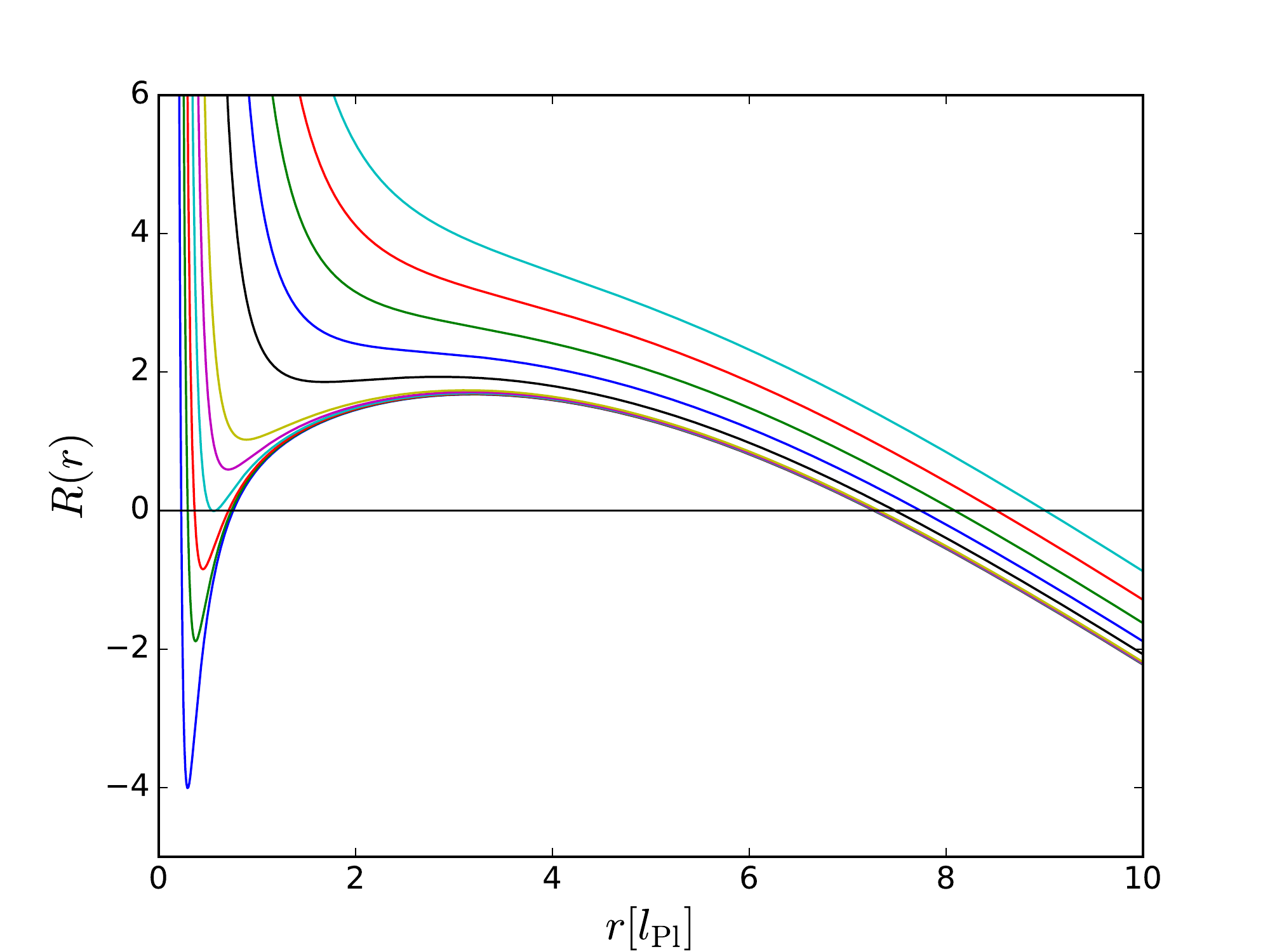}
	\caption{$R(r)$ from (\ref{Kerr_orbit}) for Kerr-AdS with $G_0=1,\Lambda_0=-0.1,M=10M_\mathrm{Pl},a=1$ and increasing particle energy $E$ from bottom to top.}
	\label{fig:kerr_ads_trajectories}
\end{figure}

\begin{figure}
	\includegraphics[width=0.49\textwidth]{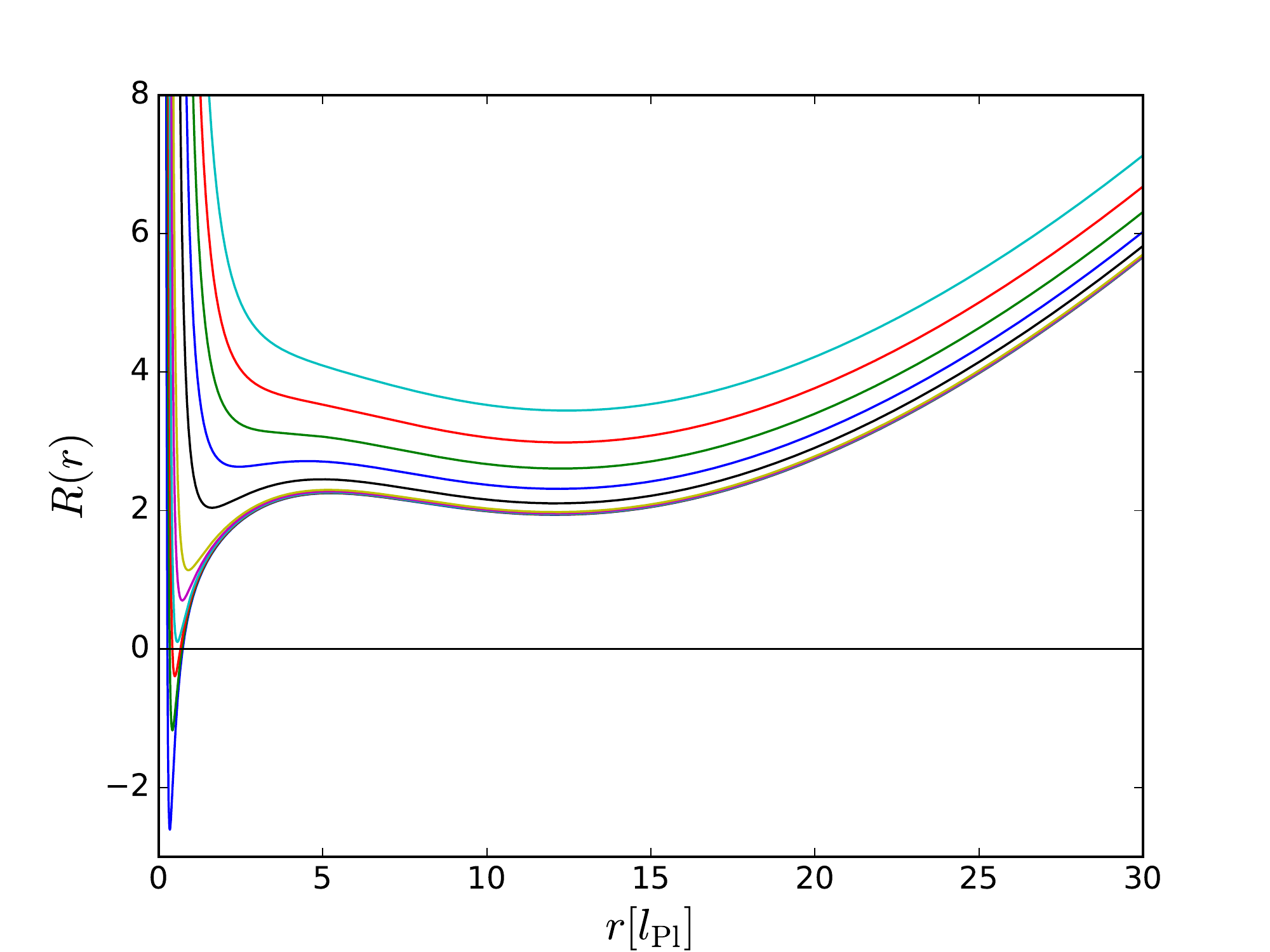}
	\caption{$R(r)$ from (\ref{Kerr_orbit}) for Kerr-dS with $G_0=1,\Lambda_0=0.01,M=10M_\mathrm{Pl},a=1$ and increasing particle energy $E$ from bottom to top.}
	\label{fig:kerr_ds_trajectories}
\end{figure}

The function $R(r)$ for Kerr-AdS is displayed in
\autoref{fig:kerr_ads_trajectories}. At large radii, the repulsiveness
of the effective AdS space-time prevents particles from escaping to
infinity. There exists a special energy $E_0$, above which observers
inevitably fall into the singularity along a terminating orbit. For
$E=E_0$, three types of orbits are possible. $R(r)$ exhibits a double
zero at $r_0$, allowing for an unstable, circular orbit. For radii
larger than $r_0$, we find a bound orbit, crossing both horizons.
Particles starting at $r<r_0$ are accelerated along terminating
trajectories and will end up in the singularity. However, if $E<E_0$,
the double root splits and we find the possibility of having bound
orbits as well as terminating ones at radii below the inner
horizon. For the smallest energies, $E\rightarrow 0$, the particle
moves from horizon to horizon.
The only difference for Kerr-dS compared to the AdS case, is that
particles can always escape to infinity, see
\autoref{fig:kerr_ds_trajectories}.

The trajectories have been calculated for an idealised, pointlike
observer, neglecting any backreaction on the geometry. However, the
location of the inner horizon is typically at about the Planck scale,
where backreaction effects should be taken into account. The quantum
improved Schwarzschild case turns out to be similar to the classical
Reissner-Nordstr\"om space-time, for which it was shown that there is
a blueshift instability at the inner (Cauchy) horizon. Additionally,
it was shown in \cite{Dafermos:2012np}, that perturbations of initial
data cause the Cauchy horizon to be replaced by a null singularity.
Due to the similarities between the quantum improved Schwarzschild and
the classical Reissner-Nordstr\"om space-time, it is tempting to
speculate that the classical findings might also hold for the quantum
case. Hence, one has to take the above results with care, especially
the many world trajectories. Summarising, there are differences
between the classical and the quantum improved geometry, but they only
become relevant at very small length scales, where the results have to
be taken with a grain of salt.

\enlargethispage{5\baselineskip}

\clearpage

\section{Curvature Singularity \& Effective Energy-momentum Tensor}
\label{sec:singularity}
Since quantum gravity effects become important in high curvature
regimes, it is expected that they alter the nature of
the curvature singularity at $r=0$. Previous results from asymptotic
safe quantum gravity \cite{Koch:2014cqa,Falls:2010he,Koch:2013owa} and
other quantum gravity scenarios, e.g. \cite{Modesto:2004xx}, predict a
substantial weakening of the singularity or even its disappearance. A
weakening of the the singularity manifests itself for instance in
changes of the Kretschmann scalar. We compute the Ricci scalar $R$ as
well as the Kretschmann scalar $K$ of the quantum improved geometries
in the UV fixed point regime, and compare the findings with the
classical result of general relativity. \autoref{curvaturescalars} 
lists the highest degree of divergence of the Ricci and Kretschmann
scalar for both investigated geometries for all discussed
matchings. Upon comparison with the classical result of general
relativity, the consistent quantum scenarios display a weakening of
the singularity but not a complete resolution.

In the quantum improved space-times, the Ricci scalar is diverging
too, because we have changed the geometry which is not a vacuum solution
of the Einstein field equations anymore. In fact, it is a geometry with an
effective energy-momentum tensor \cite{Cai:2011kd}, induced by the
running couplings. Using the classical field equations, this effective
energy-momentum tensor $T^\mathrm{eff}_{\mu\nu}$ can be computed by
calculating the Einstein tensor $G_{\mu\nu}$ from the quantum improved
metric, 
\begin{align}
G_{\mu\nu}+\Lambda_0 g_{\mu\nu}=:8\pi G_0 T^\mathrm{eff}_{\mu\nu}\;.
\label{modFE}
\end{align}
Note that $T^\mathrm{eff}_{\mu\nu}$ is covariantly conserved,
assuming a metric connection, $\nabla^\mu g_{\mu\nu}=0$, because the
Einstein tensor satisfies the Bianchi identity
$\nabla^\mu G_{\mu\nu}=0$ by construction. However, physical
interpretations of this effective energy-momentum tensor in terms of
matter have to be drawn with great care. For instance, it turns out
that the $T^\mathrm{eff}_{rr}$ is diverging at horizons, $f(r)=0$,
because $G_{rr}=\frac{f-1+rf'}{f r^2}$ and
$g_{rr}=1/f(r)$. Additionally, it has been shown in
\cite{Reuter:2010xb}, that $T^\mathrm{eff}_{\mu\nu}$ in the quantum
improved flat Kerr geometry violates the weak, the null, the strong
and the dominant energy condition. We expect similar results in the
present case, including the cosmological constant. These observations
suggest that quantum gravity contributions to the energy-momentum
tensor are of a fundamentally different nature than the ones of
conventional matter and should not be interpreted as matter. In fact,
the running couplings should be taken into account already on the
action level, resulting in different field equations. This is done for
example in Quantum Einstein Gravity (QEG) \cite{Reuter:2012id}, based
on the quantum improved Einstein Hilbert action
\begin{align}
S=\int d^4x\sqrt{-g}\left[\frac{R-2\Lambda(r)}{16\pi G(r)}\right]\;.
\end{align}
The resulting new field equations \cite{Reuter:2004nx}, based on the
runnings (\ref{analyticlambda}), read the same as (\ref{modFE}) with
\begin{align}
  8\pi G_0 T^\mathrm{eff}_{\mu\nu}=-\lambda_*k^2(r)g_{\mu\nu}+
G(r)(\nabla_\mu \nabla_\nu-g_{\mu\nu}\Box)\frac{1}{G(r)}\;.
\end{align}
It has been shown in \cite{Koch:2010nn}, that the covariant
conservation of the effective energy-momentum tensor in QEG is
equivalent to the following relation between the running couplings, 
\begin{align}
  R\,\nabla_\mu\left(\frac{1}{G(r)}\right)-2\nabla_\mu\left( 
  \frac{\Lambda(r)}{G(r)} \right)=0\;.
\label{QEGT}
\end{align}
This relation is not satisfied by our quantum improved
Schwarzschild-(A)dS and Kerr-(A)dS metrics, meaning that they are not
solutions to the new field equations (\ref{modFE}) with (\ref{QEGT}),
derived in the Einstein-Hilbert truncation of a potentially more
complicated fundamental action.

 \renewcommand{\arraystretch}{2}
\begin{table*}
	\begin{center}
		\resizebox{\textwidth}{!}{
		\begin{tabular}{|l|l|l|l||l|l|l|l|l|}
		    \hline				  & classical & cl. Kretschmann & qu. Kretschmann & linear & cl. radial path & qu. radial path & cl. geodesic & qu. geodesic\\ 
		    \hline
		    \hline
			$R_\mathrm{Sch}$ & $4\Lambda_0$ & $\sim \mathrm{const}$ & $\sim r^{-3/2}$ & $\sim r^{-2}$ & $\sim \mathrm{const}$ & $\sim r^{-2}$ & $\sim \mathrm{const}$ & $\sim r^{-3/2}$\\
			\hline
			$K_\mathrm{Sch}$ & $\sim r^{-6}$ & $\sim r^{-6}$ & $\sim r^{-3}$ & $\sim r^{-4}$ & $\sim r^{-6}$ & $\sim r^{-4}$ & $\sim r^{-6}$ & $\sim r^{-3}$ \\
			\hline
			\hline
			$R_\mathrm{Kerr}$  & $4\Lambda_0$ & $\sim r^{-3}$ & $\sim r^{-2}$ & $\sim r^{-4}$ & $\sim r^{-4}$ & $-$ & $\sim r^{-4}$ & $-$ \\ 
			\hline
			$K_\mathrm{Kerr}$ & $\sim r^{-6}$ & $\sim r^{-6}$ & $\sim r^{-4}$ & $\sim r^{-8}$ & $\sim r^{-8}$ & $-$ & $\sim r^{-8}$ & $-$\\
			\hline
		\end{tabular}}
		
	\end{center}
	\caption{Ricci scalar $R$ and Kretschmann scalar $K$ for Schwarzschild- and Kerr-AdS for different matchings compared to the classical result.}
	\label{curvaturescalars}
\end{table*}

\section{Horizon Temperatures and Black Hole Evaporation}\label{sec:temperature}
In this section, we first establish the fact, that surface gravities 
in space-times based on the quantum improved version of
the radial path proper distance are divergent, before discussing the
Hawking temperatures in the Kretschmann scenario. Finally, we will
discuss implications on the black hole evaporation process.

The Hawking temperature of a black hole in flat space received by an
observer at infinity is given by $T_\mathrm{H}=\frac{\kappa}{2\pi}$
\cite{hawking1975}, with surface gravity $\kappa$ of the event
horizon.  For an observer at finite distance $r$ in the static region
outside the black hole, the above expression is modified by a redshift
factor, 
\begin{equation}
T_\mathrm{H}=\frac{\kappa}{2\pi}\frac{1}{\sqrt{g(K,K)}}\;,
\label{t_observed}
\end{equation}
where $g(K,K)$ is the norm of the static Killing vector $K$. In more
general terms, a surface gravity can be assigned to any Killing
horizon of a space-time. Gibbons and Hawking showed in
\cite{Gibbons:1977mu}, that cosmological horizons also emit
radiation which can be detected by an observer in the static
space-time region. In general, emission is a consequence of the observer not
being able to access the space-time hidden behind the horizon(s),
thereby being fundamentally unable to measure the quantum state of the
complete universe (see \cite{Gibbons:1977mu} for a more detailed
discussion). The notion of a horizon temperature only appears to be
meaningful for observers in a static space-time region, since only
such observers detect radiation of this temperature. Taking
Reissner-Norstr\"om as example, this is only the case for the region
outside the black hole. In between the horizons, the space-time is not
static anymore and inside the inner horizon, the space-time is static
again, but connected to the singularity. This would require to impose
boundary conditions at the singularity, being far from obvious. Hence
in the following, we only refer to a horizon having a
temperature, if the horizon is the boundary of a static region, not
connected to the singularity. In appendix \ref{app:killinghorizon},
horizons in the quantum improved space-time are shown to be Killing
horizons, thus a surface gravity can be assigned to each of them.

Technically, the surface gravity $\kappa$ of a Killing horizon can be
computed by taking the covariant derivative of the norm of the Killing
vector, or alternatively via a periodicity in Euclidean time
introduced in \cite{Gibbons:1976ue}. In any case, we find
\begin{align}
 \kappa_\mathrm{Sch}=\frac{1}{2}\abs{f'(r_0)}\quad \&\quad 
 \kappa_\mathrm{Kerr}=\frac{1}{2}\frac{\abs{\Delta_r'(r_0)}}{(r_0^2+a^2)}\;,
 \label{TBH}
\end{align}
$r_0$ being the radial coordinate of the horizon. Since horizons are
zeros of $f(r)$ and $\Delta_r(r)$ respectively, (\ref{Dradial}) implies
that the derivative of the proper distance $D'(r)$ diverges at the
horizons for the quantum version of the radial path. As addressed in
appendix \ref{app:divergences} in detail, this does not necessarily
mean that the proper distance itself is diverging at a horizon, unless
the horizon is extremal. But computing the surface gravity explicitly
via (\ref{TBH}) generates the following terms, proportional to $D'(r)$,
and therefore diverging at the horizons, 
\begin{align}
f'(r)&\sim\frac{2}{3r^2} \left(-\frac{6 G_0^2\, g_*\, \lambda_*\, M\, r\, D(r)}{\left(g_* \lambda_* D^2(r)+G_0\right)^2}+\frac{r^4}{ D^3(r)}\right)D'(r)\;,\nonumber \\[2ex]
\Delta'(r)&\sim\left(\frac{2 r^2 \left(a^2+r^2\right)}{3 D^3(r)}-\frac{4 G_0^2\, g_*\lambda_*\, M r\, D(r)}{\left(g_* \lambda_* \,D^2(r)+G_0\right)^2}\right)D'(r)\;.
\end{align}
The terms in the brackets are in general non-vanishing at the
horizons. In particular, this holds also for arbitrary large masses in
the classical regime, where it is known that the surface gravity and
Hawking temperature stays finite. This is the main reason why we
consider the scale identification based on the quantum radial path as
unphysical. In contrast, along with the proper distance based on a
geodesic, the construction based on the Kretschmann scalar shows no
divergent behaviour at the horizons and therefore leads to finite
Hawking temperatures.

\begin{figure}
	\includegraphics[width=0.5\textwidth]{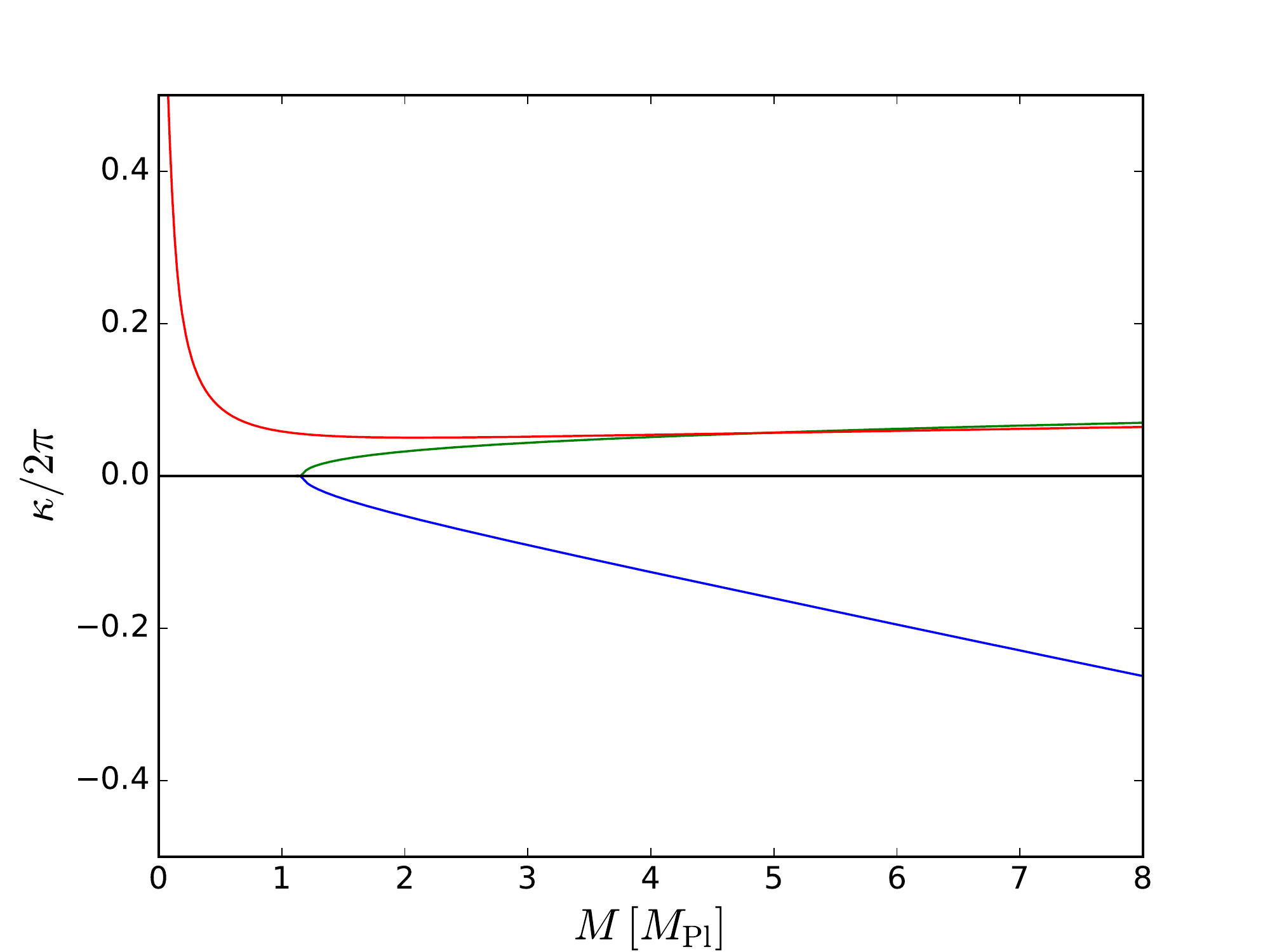}
	\caption{$f'(r)$ as function of the mass $M$ for the quantum
          improved Schwarzschild-AdS geometry for
          $\Lambda_0=-0.1$. Inner horizon in blue, outer horizon in
          green. The outer horizon agrees with the temperature of the
          event horizon in general relativity in red for large
          masses. Taking absolute values yields the surface
          gravities.}
	\label{fig:temperature_sch_AdS}
\end{figure}

\begin{figure}
	\includegraphics[width=0.5\textwidth]{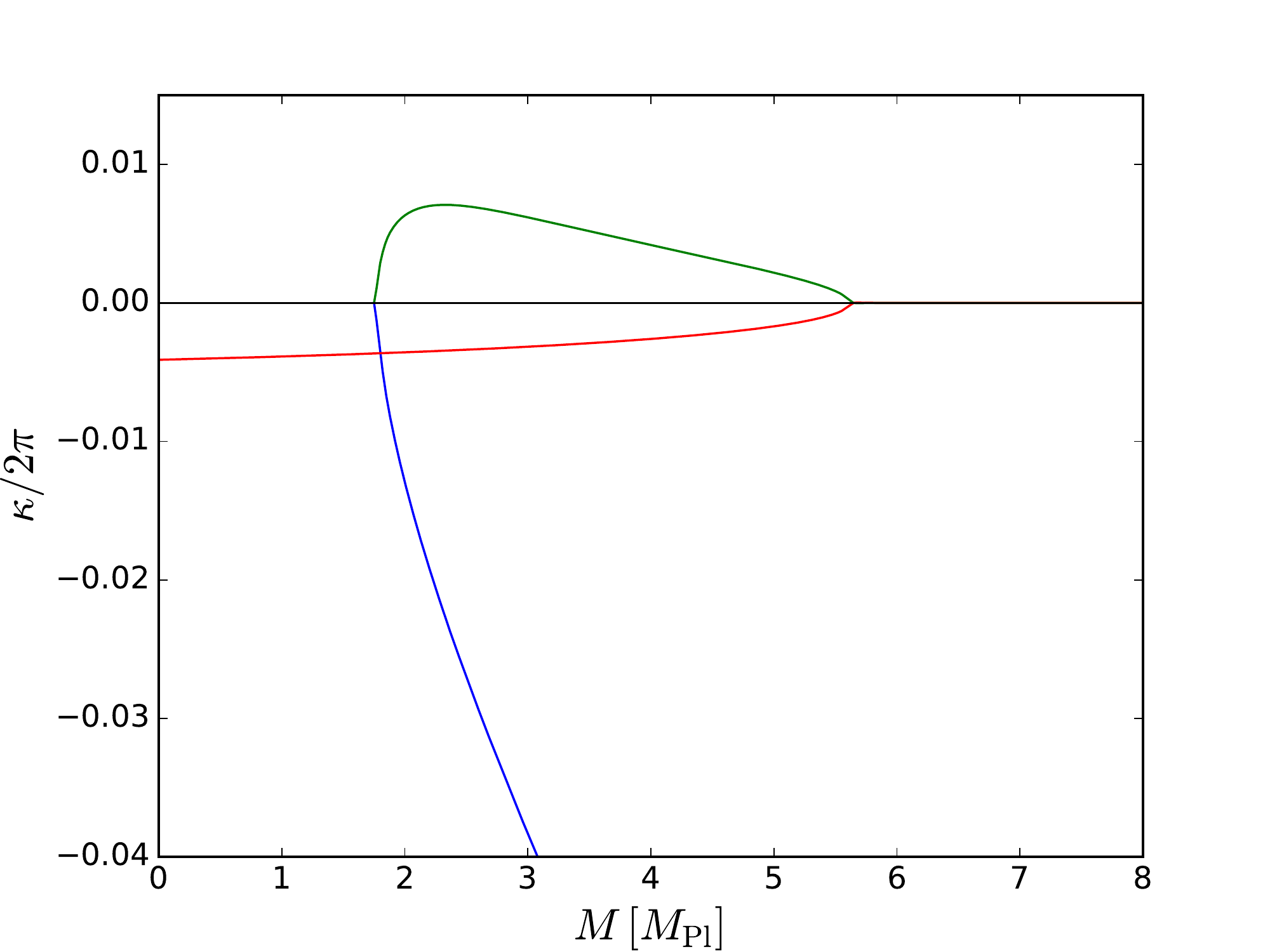}
	\caption{$f'(r)$ as function of the mass $M$ for the quantum
          improved Schwarzschild-dS geometry for
          $\Lambda_0=0.001$. The cosmological horizon in red, the
          inner black hole horizon in blue and the outer black hole
          horizon in green. Taking absolute values yields the surface
          gravities.}
	\label{fig:temperature_sch_dS}
\end{figure}

Next, we discuss the mass dependence of the surface gravities,
focusing on the quantum Kretschmann scenario from now on. It suffices
to look at the slope of the lapse function at each horizon, since it
is proportional to the surface gravity. The results for
Schwarzschild-AdS and Schwarzschild-dS can be found in
\autoref{fig:temperature_sch_AdS} and \autoref{fig:temperature_sch_dS},
the plots for the Kerr cases are qualitatively the same. The whole
evolution, appearance and disappearance of horizons is driven by the
formation of a minimum of the lapse function.  The quantum improved
Schwarzschild-AdS space-time exhibits no horizon up to the critical
mass $M_\mathrm{c}\approx1.2M_\mathrm{Pl}$. At $M=M_\mathrm{crit}$,
the minimum of the lapse function is at zero, hence the slope is zero
and so are the surface gravities. For growing mass, the slope becomes
steeper because the minimum expands, thus the surface gravities grow
in amplitude. In contrast, $\kappa_\mathrm{cl}$ in general relativity
diverges for $M\rightarrow 0$. However, the surface gravity of the
outer horizon matches the classical one for sufficiently large masses.
The Schwarzschild-dS scenario can have up to three horizons and two
special masses, $M_*\approx2M_\mathrm{Pl}$ and
$M^*\approx5.8M_\mathrm{Pl}$, at which two of the three horizons
merge. Starting in the $M<M_*$ regime, there is no black hole, but
only the cosmological horizon. The case $M=M_*$ corresponds to the
case $M=M_\mathrm{c}$ from above. For $M_*<M<M^*$, there are three
horizons and the back hole gets bigger for increasing mass, until
$M=M^*$, when the black hole has reached its maximal size and its
outer horizons merges with the cosmological horizon to an extremal
horizon with zero temperature.

In AdS space-times, an observer in the static region could only
measure a temperature coming from the black holes' event horizon,
whereas in dS space-times, the observer would measure a mixture of two
thermal spectra at different temperatures, one coming from the back hole and
one from the cosmological horizon. In the static region outside the
black hole, one valid choice for the Killing vector in
(\ref{t_observed}) is $K=\partial/\partial t$, yielding
$g(K,K)=g_{tt}$. In the Schwarzschild geometries, this implies that an
observer located at a horizon would measure an infinite temperature,
in accordance with general relativity. In Schwarzschild-AdS, the
temperature drops to zero for an infinitely distant observer, as
$g_{tt}$ diverges.\\
In the dS-scenario, there exists a distance between
the horizons, at which the observed temperature becomes minimal, s
because $f$ has a maximum. In
the Kerr geometries, $\partial/\partial t$ is timelike only outside
the ergoregion. A static Killing vector field for the entire region
outside the black hole can be obtained by linearly combining the two
Killing vectors of a Kerr space-time, see appendix
\ref{app:killinghorizon}. Since all above observations equally apply
for classical as well as quantum improved space-times, the is no
qualitative difference for an observer measuring horizon temperatures
in a classical or a quantum space-time, except in the Planckian
regime.

As final point, we would like to address the black hole evaporation
process. A standard mechanism to form black holes is gravitational
collapse. If the mass of a collapsing object is larger than the
Tolman-Oppenheimer-Volkoff mass around $2M_\odot$, no other force can
counterbalance gravity and the object collapses to form a black
hole. Assuming that a macroscopic Schwarzschild or Kerr black hole has
formed via this process, well above the critical mass, it will emit
Hawking radiation and thereby lose energy. This causes the black hole
to shrink steadily, as its mass is decreasing. This process continues,
until the critical mass $M_\mathrm{crit}$ is reached. The temperature
then becomes zero and therefore the radiation stops. Hence, the naked
singularity case with $M<M_\mathrm{crit}$ can never be reached via
this process and we end up with a zero temperature, Planck-sized,
extremal black hole, often referred to as remnant. This remnant serves
as shield, guaranteeing that the cosmic censorship conjecture remains
satisfied. However, in \cite{Aretakis:2012ei} it was shown that
extremal black hole configurations with zero temperature suffer from
an instability at the extremal horizon. Remnant endpoints were also
found in other studies within asymptotic safety
\cite{Bonanno:2006eu,Falls:2010he} and beyond
\cite{Chen:2014jwq}. Based on a classical expression for the proper
distance it has been shown in \cite{Koch:2013owa,Koch:2014cqa} that
the Schwarzschild-AdS black hole evaporates completely. A more
suitable set-up to discuss the evaporation process is given by the
dynamical Vaidya space-time, used in \cite{Bonanno:2006eu}. There, a
Planck-sized, cold remnant as endpoint has been found.

\section{Summary}\label{sec:summary}

In this work, the quantum improved Kerr-(A)dS black hole was studied
for the first time within a self-consistent scale identification
procedure. The latter is based on the Kretschmann scalar. The Kerr-(A)dS  geometry also
includes the Schwarzschild-(A)dS, as well as ordinary Schwarzschild
and Kerr space-times as special cases, by setting either the rotation
parameter $a$ or the cosmological constant $\Lambda_0$ to zero. 

Both
quantum improved geometries show the same global structure in terms of
a timelike curvature singularity at $r=0$ and the same number of
horizons. Furthermore, it has also been shown that the outer black hole
horizon corresponds to the classical black hole event horizon. The
timelike character of the singularity at $r=0$ in principle allows
particles to avoid the singularity. The quantum corrections to the
classical metric render the singularity less divergent, but none of
the studied scenarios was able to resolve it completely. However, this
singularity will always be dressed by a horizon, such that there is no
violation of the cosmic censorship conjecture.  

The horizons being
Killing horizons admit a temperature, causing the black hole to
evaporate. In the Planckian regime, however, the heat capacity of a
tiny black hole stays positive, $\frac{\partial T}{\partial M}>0$, in
contrast to the classical case. Thus, the evaporation process comes to
an end when the Hawking temperature of the black hole is zero, leaving
an extremal, cold, Planck-sized remnant, serving as cosmic
censor. This is a thermodynamically stable endpoint, because any
additional mass absorbed by the black hole will radiate away until the
temperature is again zero. It would be interesting to see what
implications for the black hole information paradox can be drawn from
the generic existence of such remnants.

\begin{acknowledgments}
  We thank Alfio Bonanno, Kevin Falls, Domenico Giulini 
and Alessia Platania for discussions. This work
is supported by EMMI and is part of and supported by the DFG
Collaborative Research Centre "SFB 1225 (ISOQUANT)" and also by the DFG Research Training Group "Models of Gravity".
\end{acknowledgments}

\appendix

\begin{figure*}
	\includegraphics[width=0.5\textwidth]{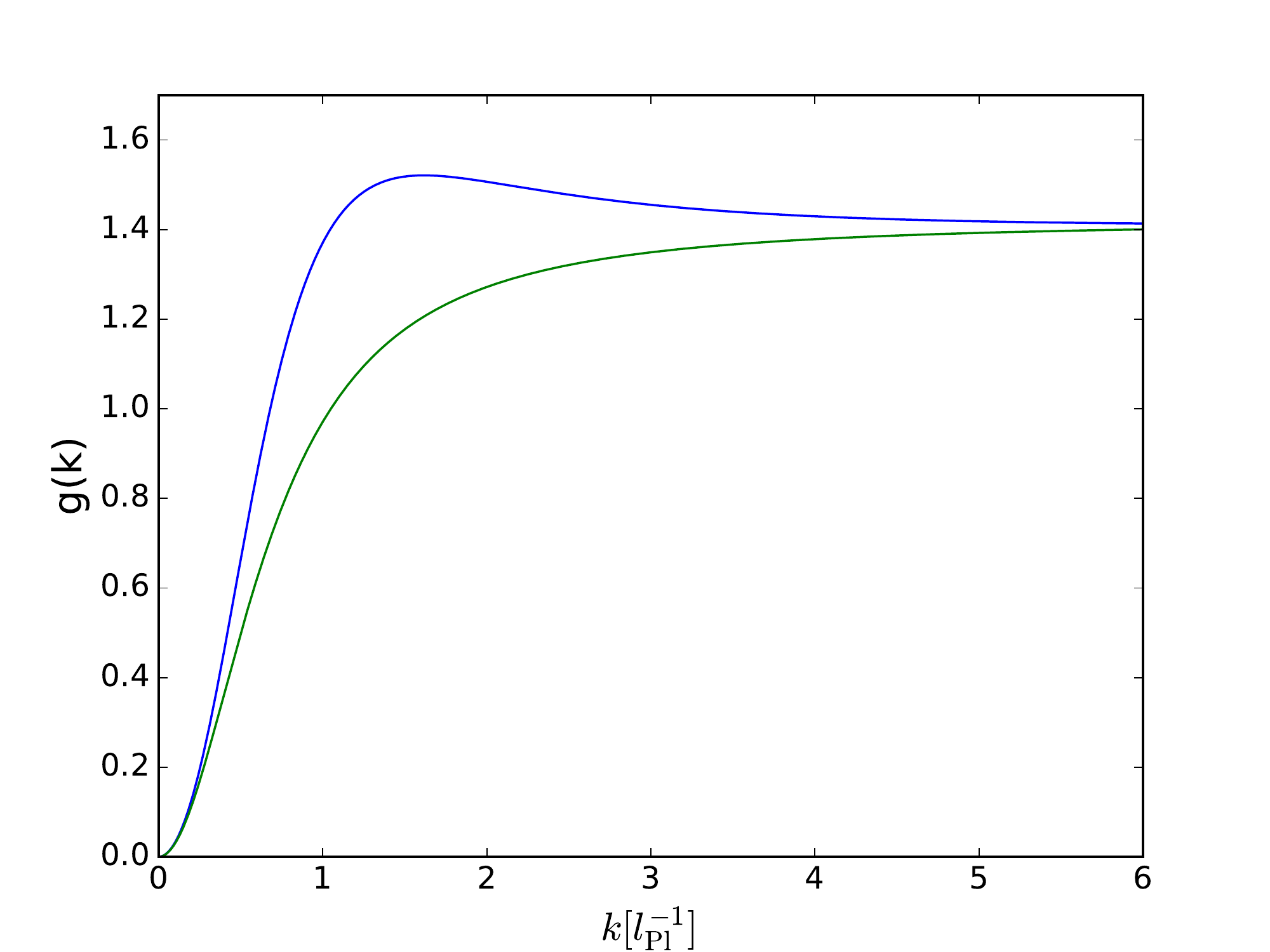}\hfill
	\includegraphics[width=0.5\textwidth]{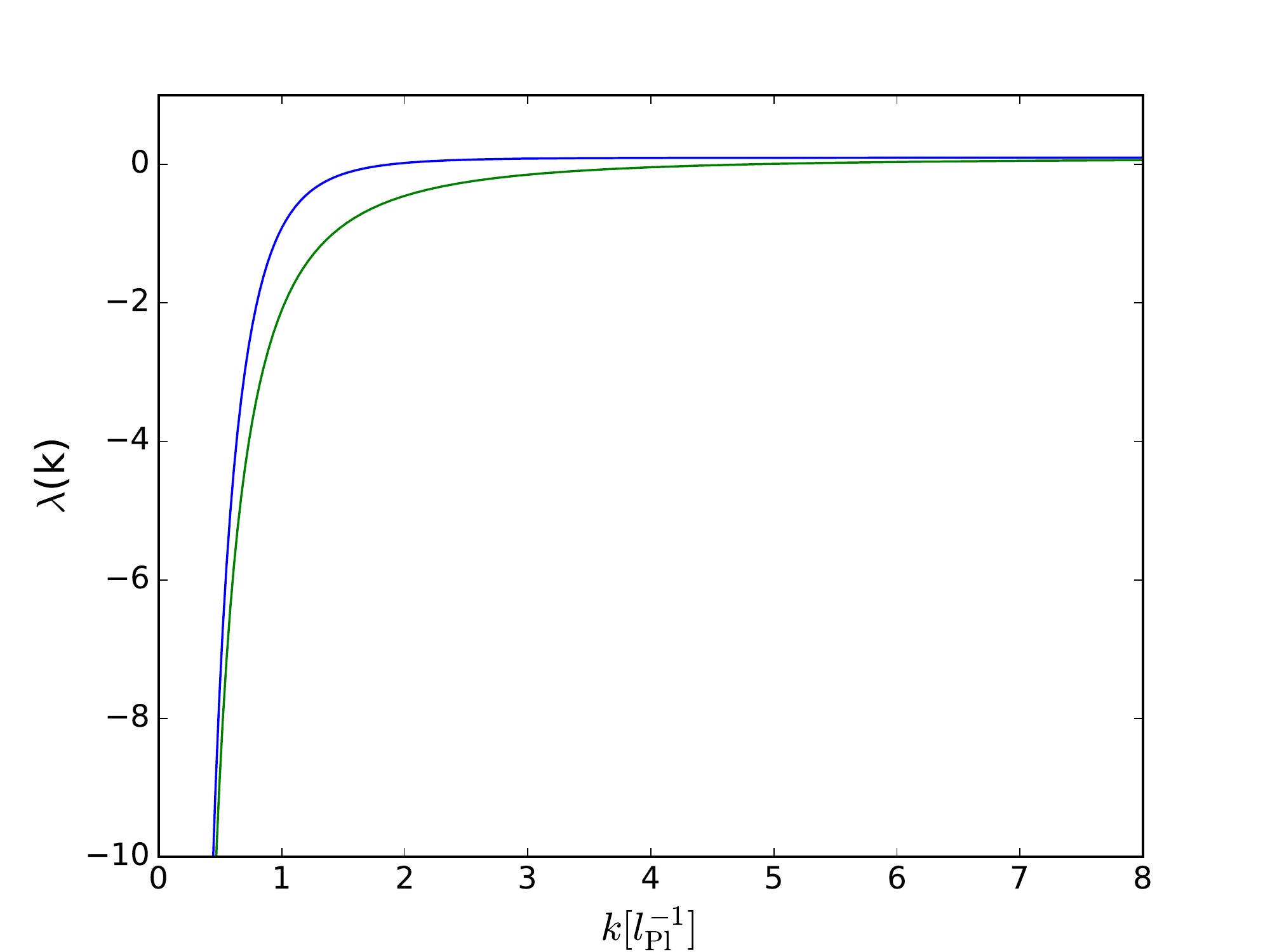}
	\caption{Running of the dimensionless couplings $g$ and
          $\lambda$ as a function of momentum scale $k$ for the
          analytical expressions from (\ref{analyticlambda}) in green
          and from a fourth order vertex expansion based on
          \cite{Denz:2016qks} in blue. Both approach their UV fixed
          point values, $g_*=1.4$ \& $\lambda_*=0.1$, for
          $k\rightarrow \infty$.}
	\label{fig:vertexexp}
\end{figure*}

\section{Choice of Scale Identification}
\label{app:choiceofxi}
Here we motivate our choice for $k(r)$ in (\ref{genscale}). Inserting
the general parametrisation $k(r)=\xi/D(r)$, into (\ref{deltar}), we
are left with
\begin{align}
\begin{split}f(r)&= 1-\frac{2M}{r}\frac{g(r)D^2(r)}{\xi^2}-\frac{r^2}{3}
\frac{\lambda(r)\xi^2}{D^2(r)}\\[2ex]\nonumber 
&\overset{\mathrm{UV}}{\underset{r\rightarrow 0}{\approx}} 1-\frac{2M}{r}
\frac{g_*D^2(r)}{\xi^2}-\frac{r^2}{3}\frac{\lambda_*\xi^2}{D^2(r)}\;, \end{split} \\[2ex]
\Delta_r &\overset{\mathrm{UV}}{\underset{r\rightarrow 0}{\approx}} (r^2+a^2)
\left(1-\frac{r^2}{3}\frac{\lambda_*\xi^2}{D^2(r)}\right)-\frac{2M}{r}\frac{g_*D^2(r)}{\xi^2} 
\label{a12}\;.
\end{align}
The numerical values of $g_*$ and $\lambda_*$ depend on the particular
RG-trajectory and parametrisation we have chosen and therefore cannot
be physical observables. However, the product $g_*\lambda_*$ is an
observable and hence independent of the particular choice of the
RG-trajectory. Its magnitude turns out to be
$g_*\lambda_*\approx 0.1$, e.g.\ in \cite{Reuter:2012id,Denz:2016qks}. In this
light, we have two choices for $\xi$ in order to make (\ref{a12})
solely dependent on $g_*\lambda_*$, 
\begin{equation}
\xi^2=g_*\qquad\text{or}\qquad \xi^2=\frac{1}{\lambda_*}\;.
\end{equation}
Thus, in (\ref{genscale}) we have chosen the second of the two
equivalent options. Varying $\xi$ for a fixed geometry
($G_0,\Lambda_0,m,a$), which is effectively done also in the quantum
Kretschmann scenario by introducing $\chi$, turns out to have only a
weak impact on the position of the inner horizon. Since it is
typically located at small radii, we recall from \autoref{table_uvlimits_D}, that varying $\xi$ mildly modifies the
UV-limit. Furthermore, we have an upper limit
$\chi<\left(3/8\right)^{1/4}$.

\section{Killing Horizons}
\label{app:killinghorizon}
In this section, we review the formal proof that every zero of
$\Delta_r(r)$ in (\ref{kerrads}) is a Killing horizon. This implies
that a constant surface gravity and thereby a temperature can be
associated to each horizon. The Schwarzschild-(A)dS case is
automatically contained by taking $a\rightarrow 0$.

Starting from the Kerr-(A)dS metric (\ref{kerrads}), assume that
$\Delta_r(r)$ has $j$ positive roots, i.e. can be written as
\begin{align}
\Delta_r(r)=\prod_{i=0}^{j}(r-r_i)\qquad\text{with}\qquad 0\le r_0\le r_1<...\le r_j\;.
\label{RR}
\end{align}
The horizons are the hypersurfaces $r=r_i=\mathrm{const}$. Since the
space-time is axisymmetric and stationary, we have two commuting
Killing vector fields: $\left(\frac{\partial}{\partial t}\right)^a$ is
stationary, at least in some region of the space-time, and
$\left(\frac{\partial}{\partial\phi}\right)^a$ manifests the symmetry
axis. We now have to construct a Killing vector field $\xi^a$, that is
normal to, and null on these horizon hypersurfaces. The most general
form for $\xi^a$ would be a linear combination of both Killing vector
fields, 
\begin{align}
\xi^a=\left(\frac{\partial}{\partial t}\right)^a+\alpha\left(\frac{
\partial}{\partial\phi}\right)^a\;,
\end{align}
with a constant $\alpha$. We will fix this constant later by requiring
that $\xi^a$ should vanish at the horizons. But first, we must change
from Boyer-Lindquist coordinates (\ref{kerrads}), to coordinates that
leave the metric regular at the horizons. Such coordinates are induced
by the principal null directions of the space-time. The Kerr-(A)dS
space-time is of algebraic type D, thus admits two distinct principal
null directions, referred to as ingoing and outgoing. They can be
represented in Boyer-Lindquist coordinates by the following vectors, 
\begin{align}
n_\pm^\mu=\left(\frac{r^2+a^2}{\Delta_r}\,\Xi\, , \pm1\, , 0\, , \frac{a}{
\Delta_r}\,\Xi \right)\;,
\end{align}
where $+1$ is outgoing and $-1$ ingoing. They now induce outgoing and
ingoing coordinates, being the Kerr-(A)dS counterparts of
Kerr-coordinates in flat space. We will select the outgoing version,
but in principle we could also work with ingoing ones.  The outgoing
Kerr-(A)dS coordinates $(v,\chi)$ are defined as, 
\begin{align}\nonumber 
\mathrm{d}v&=\mathrm{d}t+\Xi\frac{r^2+a^2}{\Delta_r}\mathrm{d}r\\[2ex]
\mathrm{d}\chi&=\mathrm{d}\phi+\Xi\frac{a}{\Delta_r}\mathrm{d}r\;.
\end{align}
Inserting these back into (\ref{kerrads}), leaves us with the metric
in terms of Kerr-(A)dS coordinates $(v,r,\theta,\chi)$,
\begin{align}\nonumber 
\mathrm{d}s^2&=-\frac{1}{\rho^2\Xi^2}\left( \Delta_r-\Delta_\theta
\, a^2\sin^2\theta\right)\mathrm{d}v^2+\frac{2}{\Xi}\mathrm{d}v\,
\mathrm{d}r\\[2ex]\nonumber 
&-\frac{2a\,\sin^2\theta}{\rho^2\Xi^2}\left( (r^2+a^2)\Delta_\theta-
\Delta_r\right)\mathrm{d}v\,
\mathrm{d}\chi\\[2ex]\nonumber 
& -\frac{2a\,\sin^2\theta}{\Xi}\mathrm{d}\chi\,\mathrm{d}r 
+\frac{\sin^2\theta}{\rho^2\Xi^2}\Bigl( \Delta_\theta(r^2+a^2)^2 \\[2ex]  
& -
\Delta_r a^2\sin^2\theta\Bigr)\mathrm{d}\chi^2+\frac{\rho^2}{\Delta_\theta}
\mathrm{d}\theta^2\;.
\label{metrickerrads}
\end{align}
One can check that (\ref{metrickerrads}) reduces to Kerr coordinates for $\Lambda=0$. The Killing vector field $\xi^a$ now reads 
\begin{align}
\xi^a=\left(\frac{\partial}{\partial v}\right)^a+\alpha\left(\frac{\partial}{\partial\chi}\right)^a\;.
\end{align}
Requiring that $\xi^a$ is null on the horizons $r=r_i$ yields
\begin{align}\nonumber 
\xi^2|_{r=r_i}&=\left[g_{vv}+2\alpha\, g_{v\chi} +\alpha^2\,g_{\chi\chi}\right]_{r=r_i}\\[2ex]
&=\frac{\Delta_\theta\sin^2\theta}{\rho_i^2\,\Xi^2}\left[a-\alpha(r_i^2+a^2)\right]^2\overset{!}{=}0
\end{align}
and therefore
\begin{align}
\alpha=\frac{a}{r_i^2+a^2}\;.
\end{align}
Thus, we have found a family of vector fields $\left(\xi^a\right)_i$,
being null at one horizon at a time. In order to show that the
hypersurfaces $r=r_i$ are Killing horizons, it remains to be checked
if $\xi^a$ is hypersurface orthogonal,
i.e. $\xi_a=\xi_\mu\mathrm{d}x^\mu\sim \mathrm{d}r$ evaluated at the
horizon,
\begin{equation}
\begin{split}
\left(\xi\right)_a|_{r=r_i}=\left[g_{\mu\nu}\xi^\nu
\mathrm{d}x^\mu\right]_{r=r_i}=\frac{1}{\Xi}\left(1-\frac{a^2
\sin^2\theta}{r_i^2+a^2}\right)\mathrm{d}r\;,
\end{split}
\end{equation}
with all other components vanishing. In summary, we are able to
construct a Killing vector field $\xi^a$ which is null on, and normal
to each horizon hypersurface $r=r_i$, and hence have shown that the
horizons corresponding to the roots of $\Delta_r$ are indeed Killing
horizons.

\section{Other Matchings}
\label{app:othermatchings}
\subsection{Linear Matching}
The simplest scaling is based on a dimensional analysis, 
\begin{align}
D_\mathrm{Lin}(r)=r\;,
\label{linear}
\end{align}
which has already been adopted for instance in
\cite{Bonanno:2000ep}. In the case of an identically vanishing
cosmological coupling, is the IR-limit of the classical proper
distance along a radial path \cite{Falls:2010he}. But this matching
does not take physical scales of the underlying space-time into
account, for instance the black hole scales given by $M$ \& $a$, or
scales induced by the gravitational or the cosmological
coupling. Nevertheless, this function already gives rise to many
phenomena observed for more complicated choices and hence can serve as
a toy model.
\begin{figure}
	\includegraphics[width=0.5\textwidth]{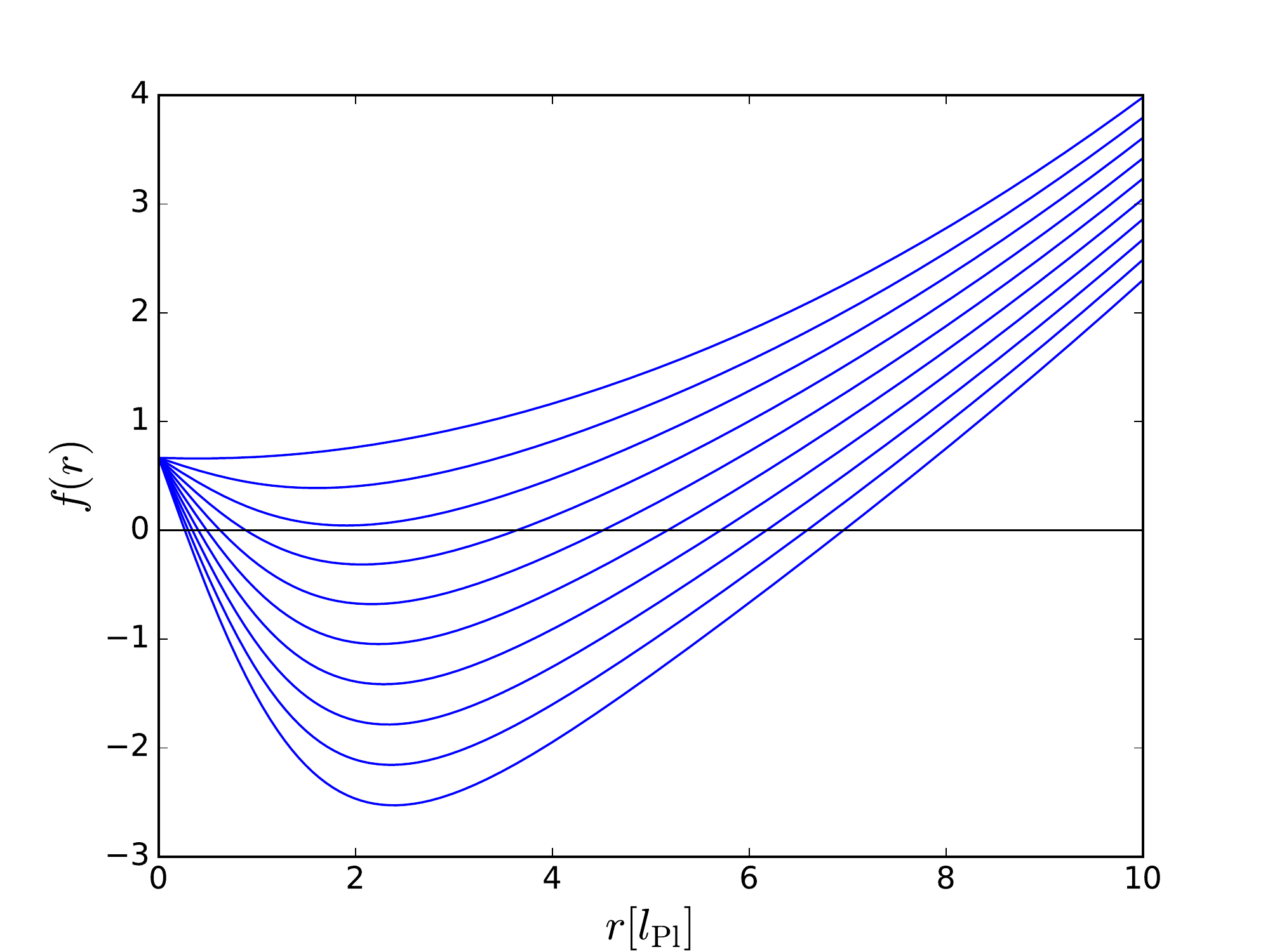}
	\caption{$f(r)$ from (\ref{schwads}) based on the linear
          matching for increasing mass from top to bottom, with
          $\Lambda_0=-0.1$,
          $M=0.1,1,2,3,4,5,6,7,8,9,10M_\mathrm{Pl}$.}
	\label{fig:fschlin}
\end{figure}
\begin{figure}
	\includegraphics[width=0.5\textwidth]{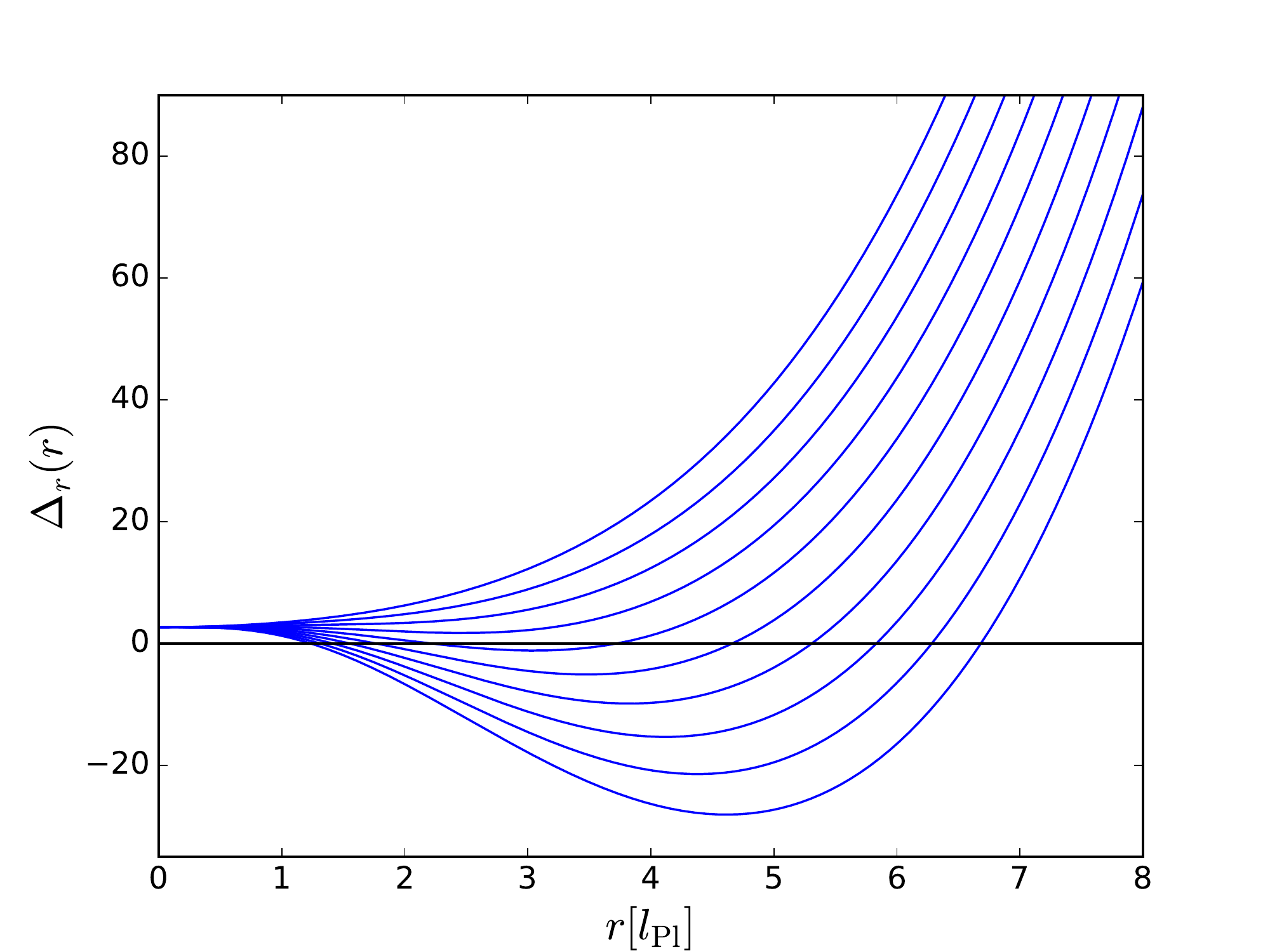}
	\caption{$\Delta_r(r)$ from (\ref{definitions}) based on the linear matching for increasing mass from top to bottom, with $\Lambda_0=-0.1$, $a=2$ and $M=0.1,1,2,3,4,5,6,7,8,9,10M_\mathrm{Pl}$.}
	\label{fig:deltalin}
\end{figure}

\subsection{Proper Distances}
\label{app:proper_distance}
We can also use the proper distance along a curve $\mathcal{C}$ in space-time 
to specify $D(r)$, 
\begin{align}
D(r)=D_\mathrm{prop}=\int_\mathcal{C}\sqrt{|g_{\mu\nu}\mathrm{d}x^\mu 
	\mathrm{d}x^\nu|}\;.
\label{properdistance}
\end{align}
This definition is diffeomorphism invariant and encodes the space-time
structure, since the gravitational and cosmological coupling typically
appear in the metric. In most cases in the literature, e.g.
\cite{Koch:2014cqa,Falls:2010he,Koch:2013owa}, the gravitational as
well as cosmological coupling have been fixed to be constants, for
instance the IR-values $\Lambda_0$ and $G_0$. However, since the
FRG-flow generically gives rise to running couplings, it is more
natural and consequent to consider this running also in the above
integral, thus $G\rightarrow G(r)$ and
$\Lambda\rightarrow\Lambda(r)$. In the following, this quantum
improvement procedure of proper distances is extended to
Schwarzschild- and Kerr-(A)dS geometries. We will provide expressions
for the proper distance along a radial path and along the geodesic of
a radially infalling observer, both for constant, as well as running
$G$ and $\Lambda$. Additionally, the UV-limit of each proper distance
is obtained, cf.~\autoref{table_uvlimits_D}.

\subsubsection{Radial Path}
Inspired by the symmetry of the space-time, we first take the
following radial path from $0$ to $r$ as integration contour
$\mathcal{C}$ in (\ref{properdistance}), 
\begin{align}
&\mathcal{C}_\mathrm{Schw-(A)dS}:\quad \mathrm{d}t=\mathrm{d}\Omega=0\;,\nonumber\\[2ex]
&\mathcal{C}_\mathrm{Kerr-(A)dS}:\quad 
\mathrm{d}t=\mathrm{d}\phi=\mathrm{d}\theta=0\quad\text{and}\quad 
\theta=\pi/2\;.
\end{align}
The restriction to the equatorial plane in the Kerr case is done for the sake 
of simplicity. Driven by the results of \cite{Reuter:2010xb} for the flat Kerr 
geometry, we assume that the varying $\theta$ will not alter our results 
qualitatively. Applying the above integration paths to (\ref{properdistance}) 
yields, 
\begin{align}
D_\mathrm{Sch}(r)&=\int_0^r \mathrm{d}\tilde{r}\, 
\sqrt{|g_{\tilde{r}\tilde{r}}|}=\int_0^r \mathrm{d}\tilde{r}\, 
\frac{1}{\sqrt{|f(\tilde{r})|}}\;,  \nonumber\\[2ex]
D_\mathrm{Kerr}(r)&=\int_0^r \mathrm{d}\tilde{r}\, 
\sqrt{|g_{\tilde{r}\tilde{r}}|}=\int_0^r 
\mathrm{d}\tilde{r}\,\sqrt{\frac{\tilde{r}^2}{|\Delta_r(\tilde{r})|}}\;,
\label{Dradial}
\end{align}
with the lapse functions
\begin{align}
f(r)&=1-\frac{2  G M}{r}-\frac{\Lambda}{3}r^2\qquad\text{and}\nonumber\\[2ex]
\Delta_r(r)&=(r^2+a^2)\left(1-\frac{\Lambda}{3}r^2\right)-2MG r\label{deltar}\;.
\end{align}
In the following, this scenario with constant $G$ and $\Lambda$ will
be referred to as \textit{classical radial path}, because the
space-time underlying the integral is a classical black hole geometry
with a cosmological
constant. 

Alternatively, we account for the running of the couplings already in
the proper distance, referred to as \textit{quantum radial path} with
$G=G(r)$ and $\Lambda=\Lambda(r)$ in the above integrals. This turns
(\ref{Dradial}) into integral equations for $D(r)$, which can be
transformed into a differential equation by taking a derivative with
respect to $r$. One can then easily see that the derivative of $D(r)$
diverges at every horizon, where $f(r)$ and $\Delta(r)$ vanish. Using
the fixed point behaviour of $G$ and $\Lambda$ in the UV, these
differential equations read for small $r$, 
\begin{align}
D_\mathrm{sch,qu}'(r)&=\frac{1}{\sqrt{\abs{1-2Mg_*\lambda_*	\frac{D_\mathrm{sch,qu}^2(r)}{r}-	\frac{r^2}{3D_\mathrm{sch,qu}^2(r)}}}}\;,\nonumber\\[2ex]
D_\mathrm{kerr,qu}'(r)&=\frac{1}{\sqrt{\abs{1+\frac{a^2}{r^2}-\frac{r^2}{3D_\mathrm{kerr,qu}^2(r)}-\frac{a^2}{3D_\mathrm{kerr,qu}^2(r)}}}}
\;.
\label{deqkerr}
\end{align}
Both classical matchings as well as the one for the quantum
Schwarzschild scenario monotonously increase and satisfy
$D(r\rightarrow 0)=0$, as can be seen from the numerical results in
\autoref{fig:Dsch_rad}. In contrast, the proper distance is
identically zero in the quantum Kerr scenario, see (\ref{epsilonequ}).

It turns out (cf.\ \autoref{sec:temperature}), that the expression for the Hawking 
temperature in a quantum improved space-time contains terms proportional to the 
derivative of $D(r)$, hence using the above construction for the proper 
distance leads to diverging Hawking temperatures at all horizons. Therefore, in 
the following we also discuss the proper distance induced by the eigentime of a 
radially infalling observer, where this feature is absent.

\subsubsection{Radial Timelike Geodesic}
The eigentime $\tau$ of an observer, initially at rest at $R$ and 
falling along a radial timelike geodesic into the singularity, can also be used to 
identify the momentum cut-off scale with a length scale by setting 
$D(r)=\tau(r)$. 
Derived in appendix \ref{app:geodesicSchwarzschild}, the eigentime for the 
Schwarzschild-(A)dS 
scenario reads
\begin{align}
D(R)=\int_0^R\mathrm{d}r\, \frac{1}{\sqrt{|E^2-f(r)|}}\;,
\label{schwgeod}
\end{align}
with $E=f(R)$ for an observer initially starting at rest. It is worth
noting that for $E=0$, the integral reduces to the one in
(\ref{Dradial}). By fixing $E$, we equivalently specify the maximal
distance $R$ of the observer from the origin. Independent on the
particular value of $E$, the proper distance again exhibits poles if
$E^2-f(r)=0$, now shifted by $E^2$ away from the horizons. Once more,
(\ref{schwgeod}) gives rise to two different proper distances,
referred to as either \textit{classical} or \textit{quantum geodesic},
depending on whether the underlying space-time is based on the
constant or
running versions of $G$ and $\Lambda$. 
ss
The analogous expression for the proper distance induced by an radial
geodesic in the Kerr-(A)dS scenario reads (see appendix
\ref{app:geodesicKerr})
\begin{align}
\begin{split}
D(R)&= \int_0^R\mathrm{d}r\, \frac{r^2}{\sqrt{|E^2\,\Xi^2\left[ 
		(r^2+a^2)^2-a^2\Delta_{r} 
		\right]-r^2\Delta_{r}|}}\;,\label{kerrgeodesic}\\[2ex]
E^2&=E^2(R)= 
\frac{R^2\Delta_R}{\Xi^2\left[(R^2+a^2)^2-a^2\Delta_R\right]}\;,
\end{split}
\end{align}
and reduces to (\ref{Dradial}) for $E=0$. Again, we achieved that
there are no poles at the horizons. Once more, we have two versions
depending on whether $G$ and $\Lambda$ are running or not. The
numerical results can be found in \autoref{fig:Dsch_geo}, however,
the proper distance in the quantum Kerr scenario is again identically
zero.

\begin{figure}
  \includegraphics[width=0.5\textwidth]{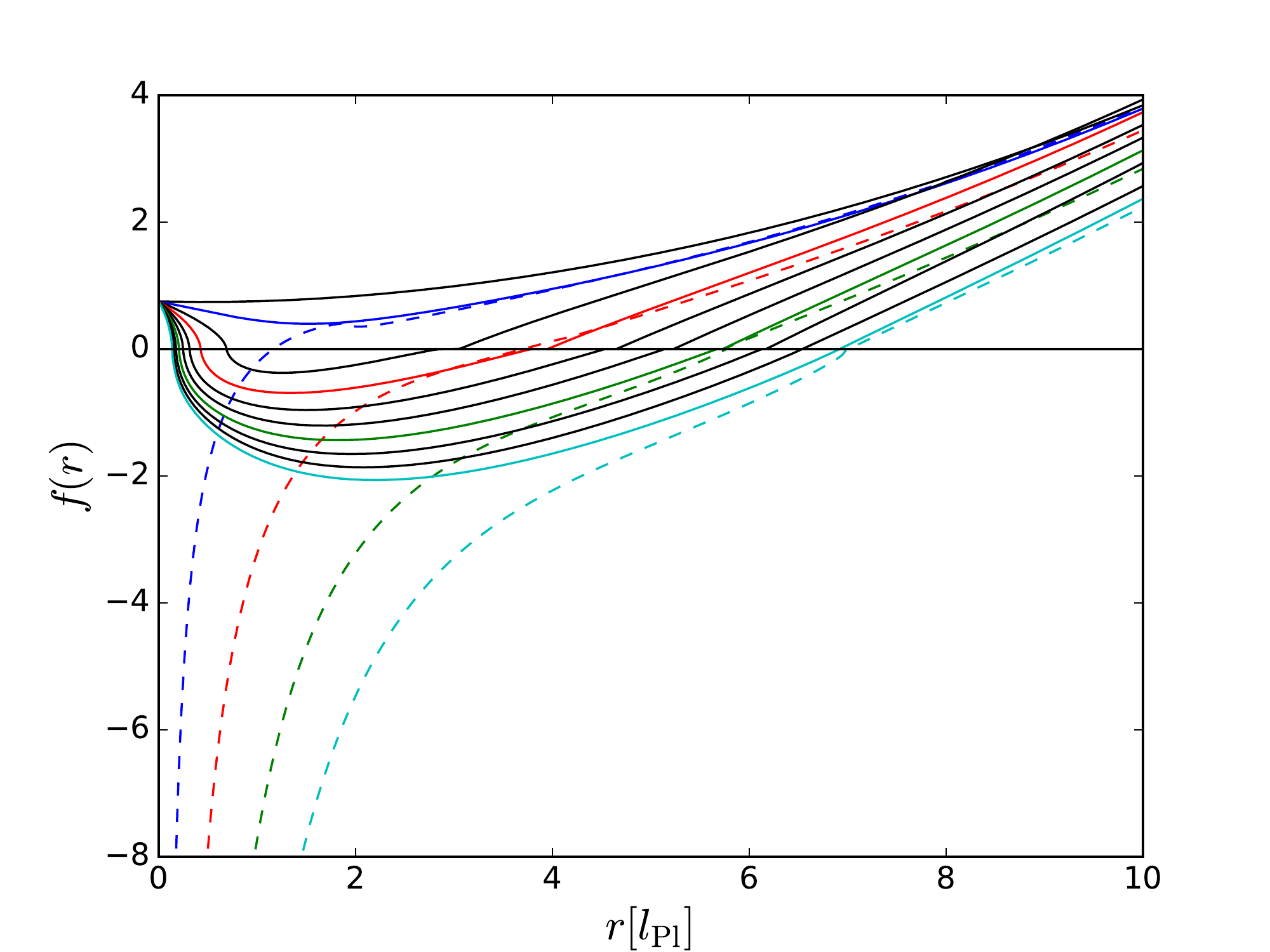}
  \caption{$f(r)$ based on the radial path matching for increasing
    mass from top to bottom. Results, where $D(r)$ is computed
    consistently in a quantum improved space-time, are shown in solid,
    the dashed curves are the ones with a classically computed
    $D(r)$. With parameters $\Lambda_0=-0.1$ and
    $M=0.1,1,2,3,4,5,6,7,8,9,10M_\mathrm{Pl}$. Curves of the same mass
    have the same colour.}
	\label{fig:fsch_rad}
\end{figure}
\begin{figure}
	\includegraphics[width=0.5\textwidth]{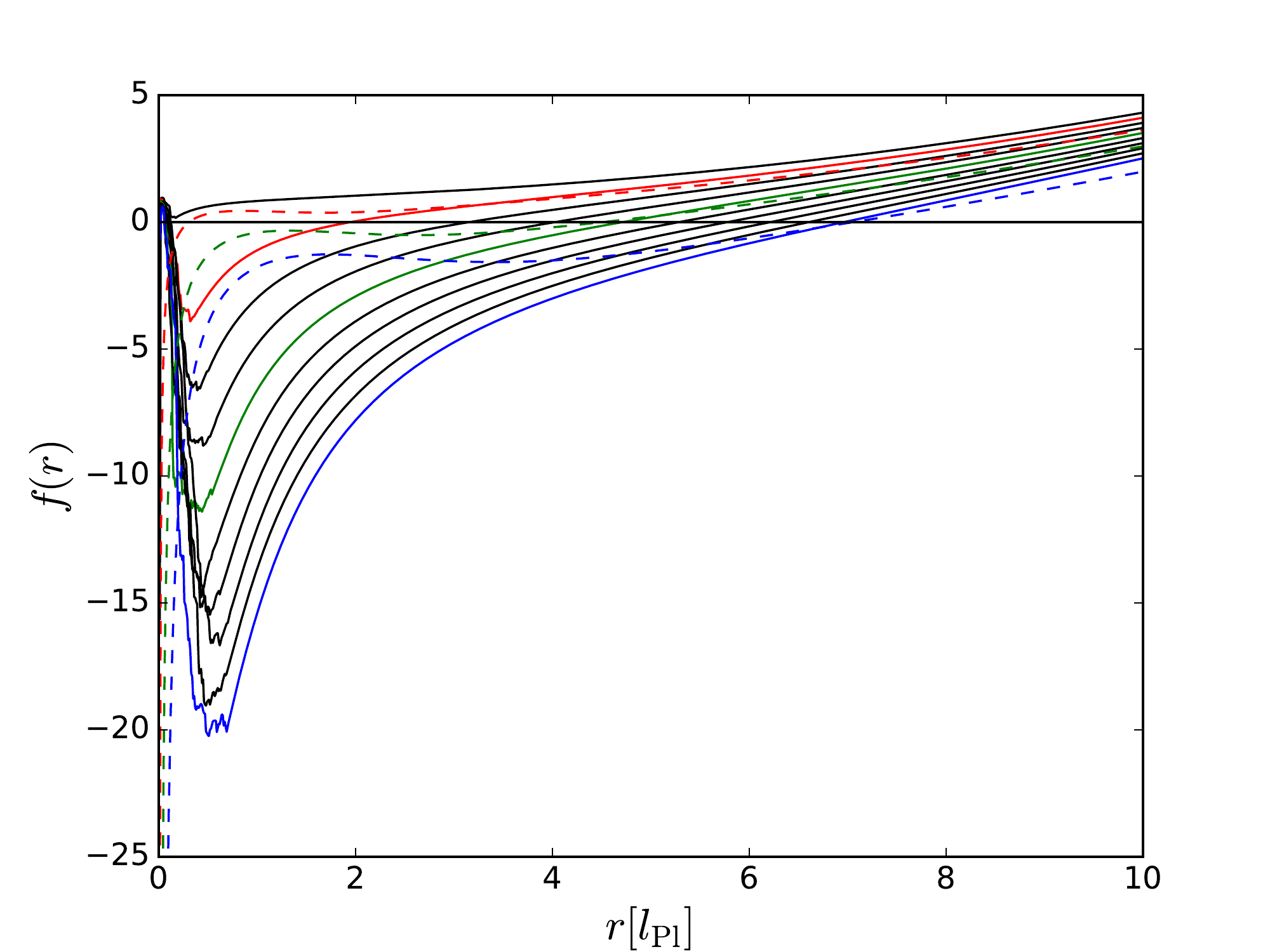}
	\caption{$f(r)$ based on the radial geodesic matching for
          increasing mass from top to bottom. Results, where $D(r)$ is
          computed consistently in a quantum improved space-time, are
          shown in solid, the dashed curves are the ones with a
          classically computed $D(r)$. With parameters
          $\Lambda_0=-0.1$ and
          $M=0.1,1,2,3,4,5,6,7,8,9,10M_\mathrm{Pl}$. Curves of the
          same mass have the same colour.}
	\label{fig:fsch_geo}
\end{figure}

\begin{table*}[]
	\resizebox{\textwidth}{!}{
		\begin{tabular}{c||l|l|l|l|l|l }
			\hline
			&\multicolumn{2}{c|}{Kretschmann}&\multicolumn{2}{c|}{Radial Path}&\multicolumn{2}{c}{Geodesic Path} \\
			&classic & quantum & classic & quantum & classic & quantum \\
			\hline\hline
			Schwarzschild& \multirow{2}{*}{$\frac{1}{3^{1/4}\,2\chi\sqrt{MG_0}}r^{3/2}$} &\multirow{2}{*}{$\left(\frac{\chi^{-4}-8/3}{48M^2(g_*\lambda_*)^2}\right)^{1/8} r^{3/4}$} & $\frac{2}{3\sqrt{2G_0M}}r^{3/2}$ & $\frac{2}{\sqrt{3}}r$ &
			$\frac{\pi}{2\sqrt{2G_0M}}r^{3/2}$ & 	$\left(\frac{67}{18Mg_*\lambda_*}\right)^{1/4}r^{3/4}$\\
			\cline{4-7} Kerr&  &  & $\frac{r^2}{2a}$ & $0$ & $\frac{\pi}{4a}r^2$ & $0$ \\
			\hline
		\end{tabular}}
		\caption{UV-limits ($r\rightarrow 0$) of $D(r)$ for all investigated matchings.}
		\label{table_uvlimits_D}
	\end{table*}

\section{Shape and Divergences of Proper Distances}
\label{app:divergences}

As can be seen from \autoref{fig:Dsch_rad}-\ref{fig:Dkretschmann}, all
functions $D(r)$ are monotonously increasing, some proper distances
display a rapid increase. In order to understand these jumps and
possible divergences, we have to look at the integral expressions for
each proper distance (\ref{Dradial}), (\ref{schwgeod}), and
(\ref{kerrgeodesic}). The expression $h(r)$ under each square root can
become zero, and if $h(r)$ has just a single root at $r=r_0<R$, the
corresponding pole is integrable, causing a jump in the proper
distance, 
\begin{align}
\begin{split}
  D(R)&=\int_0^R \mathrm{d}r\frac{1}{\sqrt{h(r)}}=\int_0^R \mathrm{d}r
  \frac{1}{\sqrt{(r-r_0)\tilde{h}(r)}}\\[2ex] 
 &\sim \int_0^R
  \mathrm{d}r\left(r- r_0\right)^{-1/2}\;,
\end{split}
\end{align}
where $\tilde{h}(r)$ has no root at $r=r_0$. However, once the
multiplicity of $r_0$ is larger than one, the pole is not integrable
anymore and $D(r)$ exhibits a divergence at $r=r_0$. In any case,
$D'(r)$ is diverging for the radial path proper distances, even at
integrable poles of $D(r)$, as can be seen from (\ref{deqkerr}). In
case of the classical radial path, the position of these poles has no
direct physical significance, however in the quantum case, the poles
are located precisely at the horizons, because then, the function
$h(r)$ is nothing other than the horizon condition. Thus, for extremal
black holes when at least two horizons coincide, the quantum proper
distance along a radial path is ill-defined. $D'(r)$ is always
diverging at the horizons leading to a diverging Hawking temperature
of the horizon, as is shown in \autoref{sec:temperature}.

For this reason, we introduce the scenario with an infalling observer
along a timelike, radial geodesic, in order to remove these problems,
only due to the poor choice of the function $h(r)$ and absent in all
other scenarios. However, it turns out, that in both proper distance
scenarios for Kerr-(A)dS with an underlying quantum space-time, the
proper distance must vanish identically, in order to satisfy the
condition $D(0)=0$. For instance, this can be seen by solving
(\ref{deqkerr}) in the limit $r\rightarrow 0$, satisfying the boundary
condition $D(0)=\epsilon$, yielding
\begin{align}
  D^\mathrm{kerr}_\mathrm{rad,UV}(r)=\frac{\epsilon}{a^{\sqrt{3}}}\left(
    \sqrt{r^2+a^2}+r\right)^{
    \sqrt{3}}\;.
\label{epsilonequ}
\end{align}
Therefore, the solution vanishes identically in the limit
$\epsilon\rightarrow 0$, which is confirmed also for the full,
numerical solution of (\ref{Dradial}). The same behaviour is found for
Kerr-(A)dS, when the scale matching is based on the geodesic in a
quantum-improved space-time.

\section{UV-limits of D(r)}
\label{app:UVlimits}
For statements about the curvature near the singularity and also for
the construction of the Penrose diagrams, the UV-limit for each proper
distance is needed.

The leading order behaviour in the UV for the classical proper
distances, i.e. constant $G_0$ and $\Lambda_0$, can be obtained from
(\ref{Dradial}), (\ref{schwgeod}) and (\ref{kerrgeodesic}) by
approximating the integral in the limit $r\rightarrow 0$.  For the
identification based on the classical Kretschmann scalar
(\ref{kretschmannident}), the UV-behaviour can easily be read off from
(\ref{kret}).

In the quantum versions, the leading order of the proper distance in
the UV-limit can be obtained by assuming a power law behaviour of the
form $D(r)=A\, r^\alpha$, with constants $A>0$ and $\alpha>0$ in order
to satisfy the boundary condition $D(0)=0$. The constants $A$ and
$\alpha$ can be determined by inserting this ansatz back into the
above equations, now being an integral, differential or functional
equation respectively. All scenarios display monotonously increasing
functions satisfying $D(0)=0$, apart from the quantum proper distance
expressions for Kerr. They are identically zero, as an iterative
algorithm for solving the integral equations shows.

\begin{figure*}
	\includegraphics[width=0.48\textwidth]{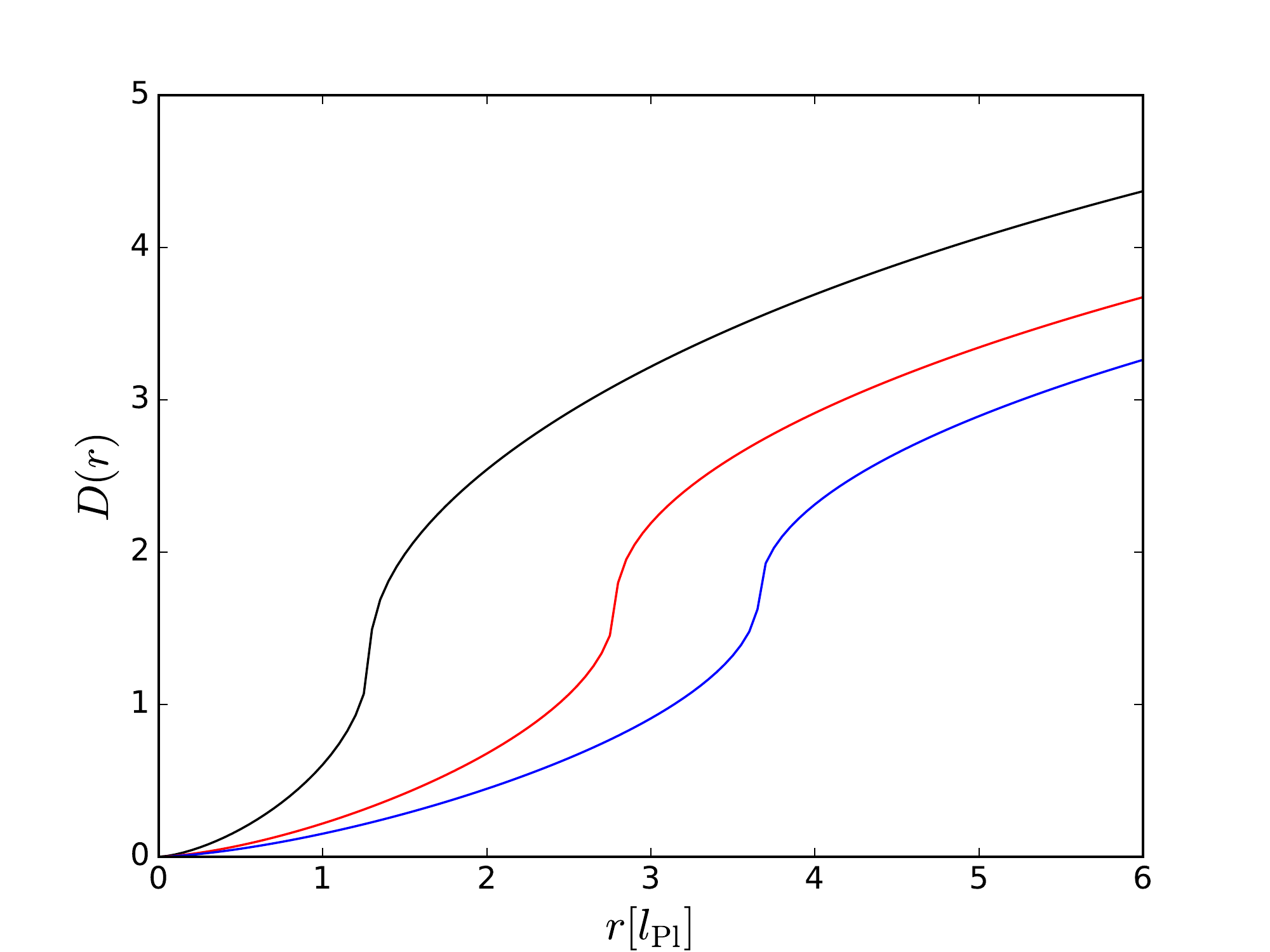}
	\hfill
	\includegraphics[width=0.48\textwidth]{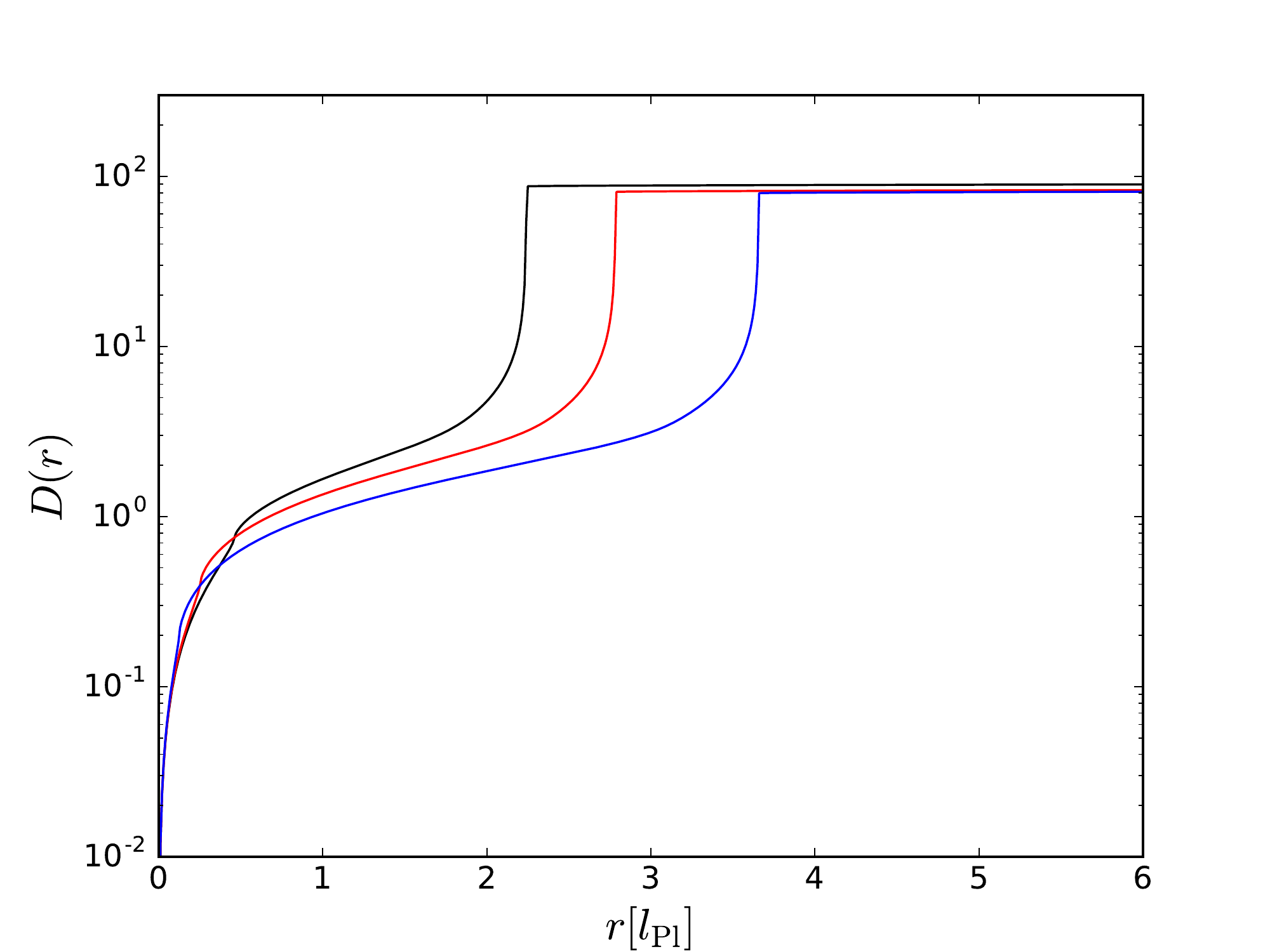}
	\caption{Left: proper distance along a radial path through a classical Schwarzschild-AdS space-time for three different masses $M=1,5,10\,M_\mathrm{Pl}$. Right: the same for a quantum Schwarzschild-AdS space-time.}
	\label{fig:Dsch_rad}
\end{figure*}

\begin{figure*}
	\includegraphics[width=0.48\textwidth]{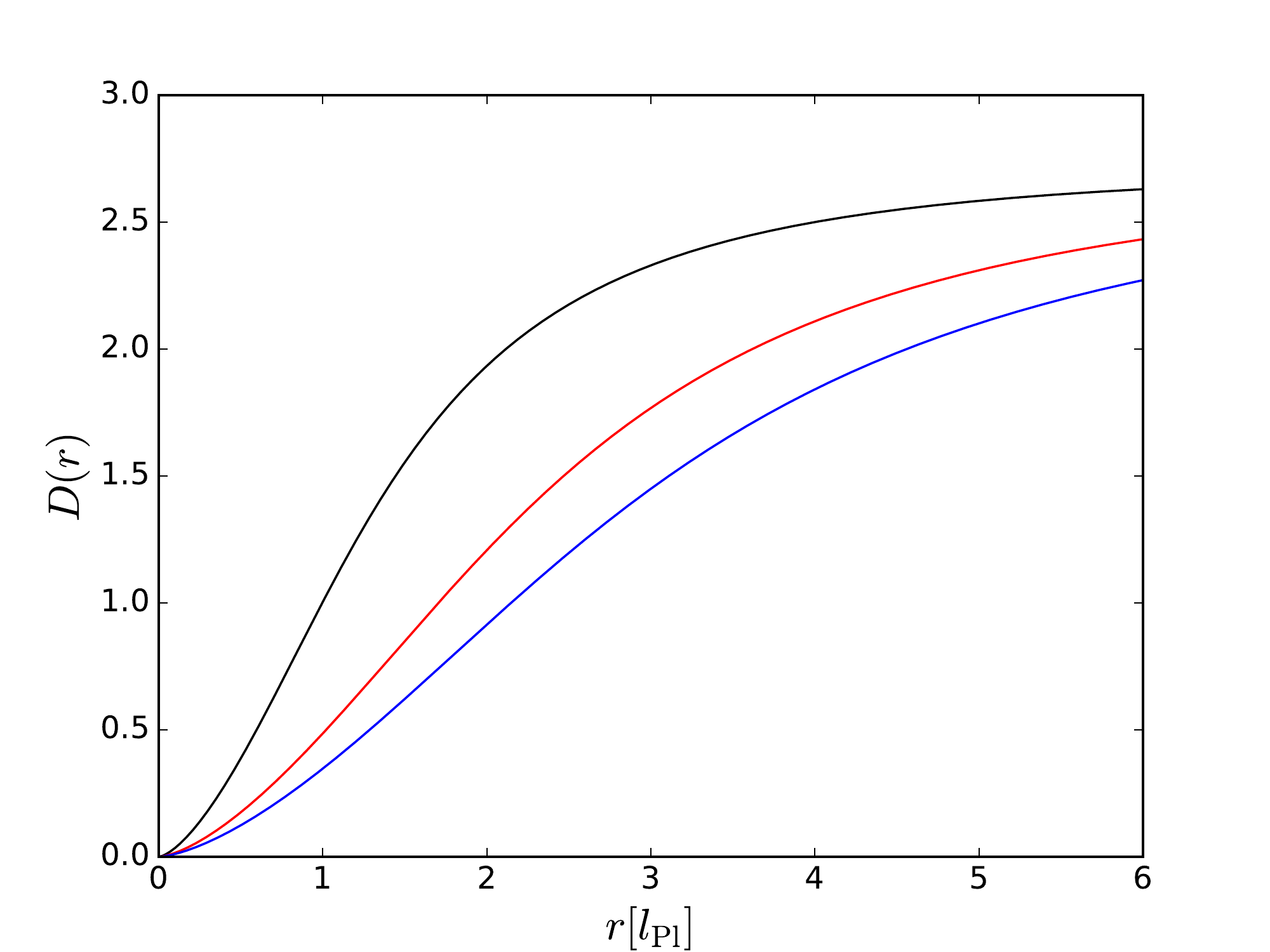}
	\hfill
	\includegraphics[width=0.48\textwidth]{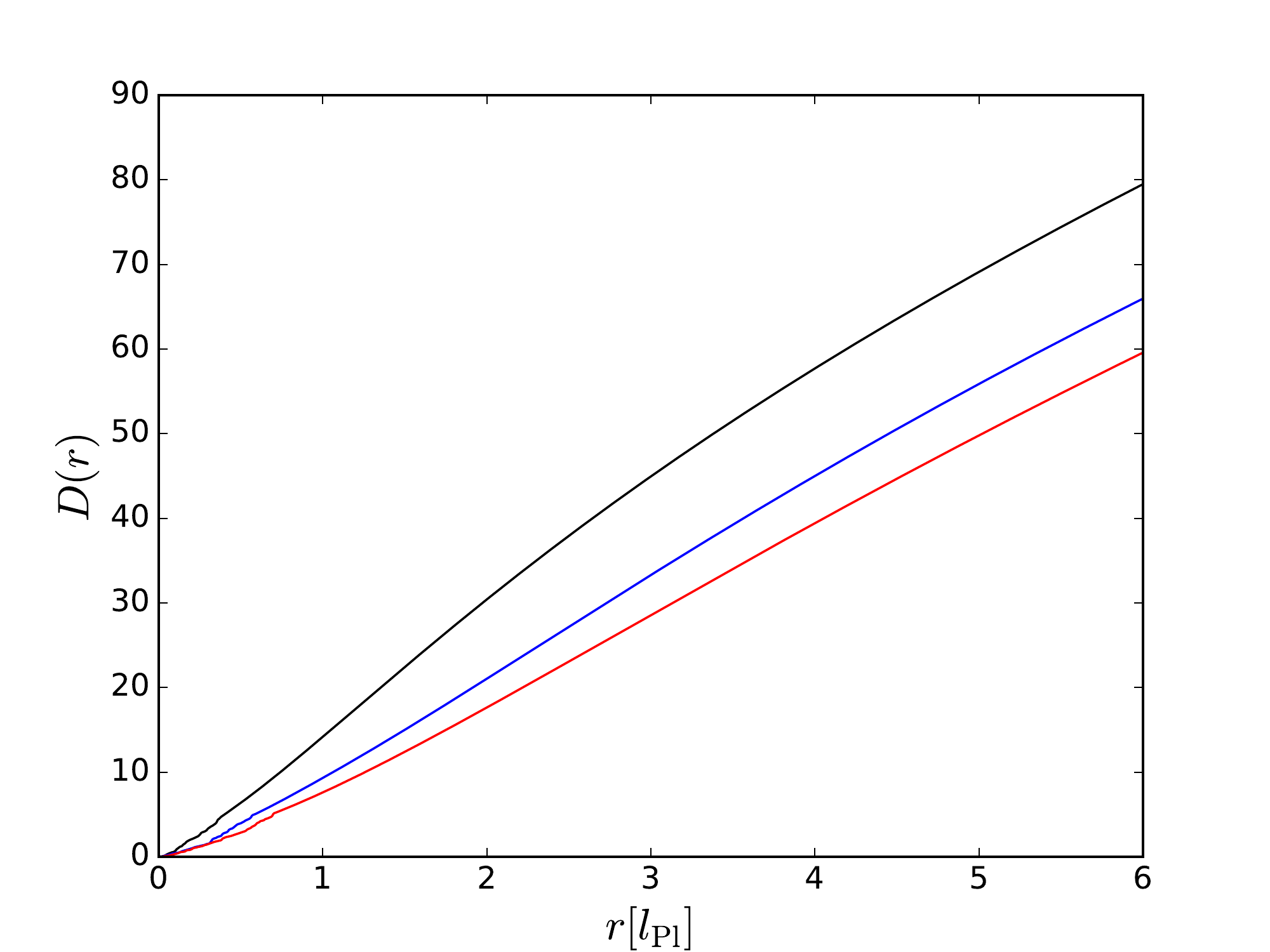}
	\caption{Left: proper distance along a radial geodesic through a classical Schwarzschild-AdS space-time for three different masses $M=1,5,10\,M_\mathrm{Pl}$. Right: the same for a quantum Schwarzschild-AdS space-time.}
	\label{fig:Dsch_geo}
\end{figure*}

\begin{figure*}
	\includegraphics[width=0.48\textwidth]{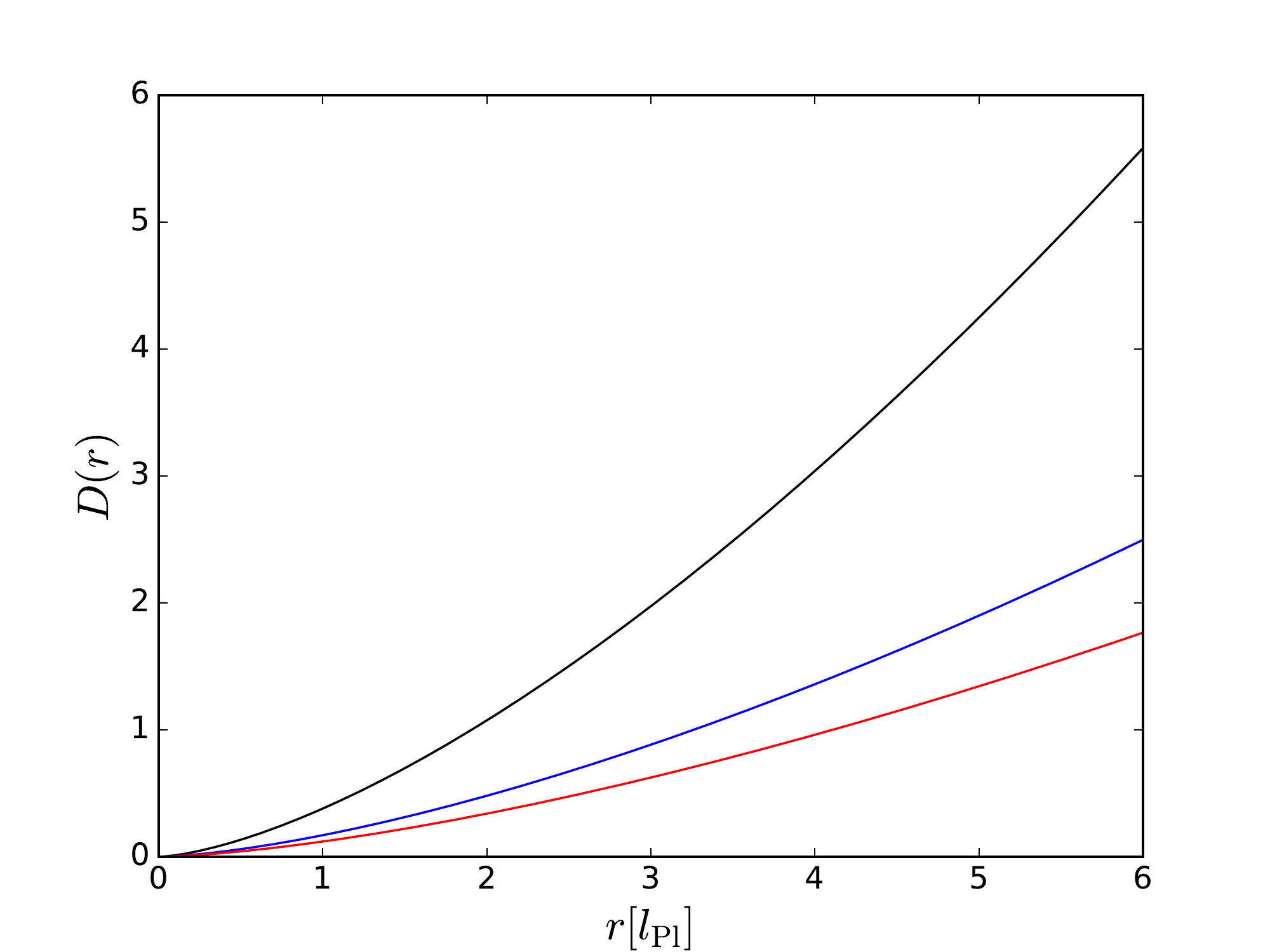}
	\hfill
	\includegraphics[width=0.48\textwidth]{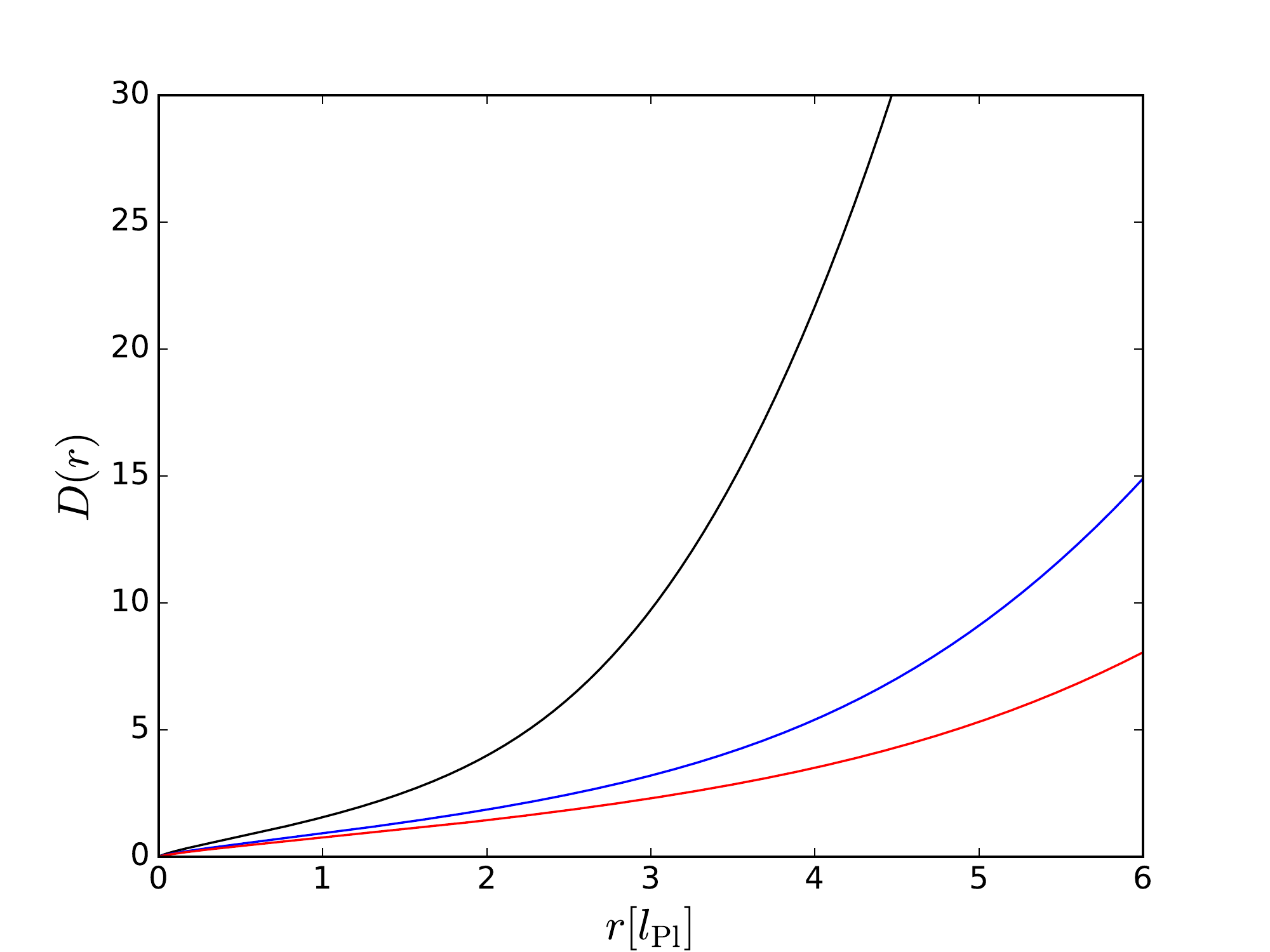}
	\caption{Function $D(r)$ in classical Kretschmann matching (left) and for quantum Kretschmann scenario (right) for three different masses $M=1,5,10\,M_\mathrm{Pl}$.}
	\label{fig:Dkretschmann}
\end{figure*}

For each scenario, the analytical UV-expression is listed in 
\autoref{table_uvlimits_D}. The numerical results for $D(r)$ are shown in
\autoref{fig:Dsch_rad}, \autoref{fig:Dsch_geo}, 
\autoref{fig:Dkretschmann}. Furthermore, the leading order exponent
$\alpha$ can be extracted numerically from the slope of the linear
relation between the proper distance $D(r)=A\,r^\alpha$ and its
integral function $\mathcal{D}(r)=\frac{A}{\alpha+1}r^{\alpha+1}$:
\begin{align}
\frac{\mathcal{D}(r)}{D(r)}=\frac{r}{\alpha+1}\;.
\end{align}
This cross-check confirms agreement between numerical exponent and the
one found analytically in \autoref{table_uvlimits_D}.

\enlargethispage{5\baselineskip}

\newpage
\section{Eigentime of an Inflating Observer in a Schwarzschild-(A)dS 
	Geometry}\label{app:geodesicSchwarzschild}
Another physically well-motivated choice for the integration path in 
(\ref{properdistance}) is the curve determined by an observer some distance 
away from the black hole, falling into the black hole along a radial timelike 
geodesic. Because the observer's four-velocity $u^a$ is conserved along 
geodesics, we normalise it to be
\begin{align}
-1\overset{!}{=}u_a u^a\;.
\end{align}
Furthermore, we can choose the coordinate system such that the motion takes 
place only in the equatorial plane $\theta=\pi/2$. Using (\ref{schwads}), the 
normalisation condition of the four-velocity in the equatorial plane reads:
\begin{align}
-f(r)\,\dot{t}^2+\frac{\dot{r}^2}{f(r)}+r^2\dot\phi^2 =-1\;,
\label{4vel}
\end{align}
where $\dot{(\;)}$ denotes the derivative with respect to the eigentime $\tau$. 
We have also conserved quantities $E$ and $L$ corresponding to the Killing 
vector fields $\xi^a=\left(\frac{\partial}{\partial t}\right)^a$ and 
$\psi^a=\left(\frac{\partial}{\partial \phi}\right)^a$:
\begin{align}
E&=-g_{ab}\xi^a u^b=f(r)\dot{t}\;,\\[2ex]
L&= g_{ab}\psi^a u^b= r^2 \dot\phi\;.
\end{align}
However, for simplicity, we will choose an observer with $L=0$. Inserting 
$E$ and $L$ back into (\ref{4vel}) to eliminate $\dot t$ and $\dot\phi$ leaves 
us with
\begin{align}
E^2=\dot{r}^2+f(r)\;.
\label{geodsch}
\end{align}
This is a type of energy equation for the observer, at least in asymptotically 
flat spacetimes. We now have to specify the initial conditions for the 
observer. In the asymptotically flat spacetime, one usually places the observer 
initially at rest at $r=\infty$, still leaving $E$ finite. However, we cannot 
do 
that in the case of a non-vanishing cosmological constant, because $f(r)$ is 
diverging for $r\rightarrow\infty$. Therefore, we take rather an observer at 
rest ($\dot{r}^2=0$) at some finite distance $R$ to determine $E$:
\begin{align}
E^2=f(R)\;.
\end{align}
The proper distance is then given by the eigentime the observer needs to arrive 
at $r=0$ after starting at $R$, i.e the integral over the eigentime along the 
geodesic:
\begin{align}
D(R)=\int_0^R\mathrm{d}r\, \frac{1}{\sqrt{|E^2-f(r)|}}=\int_0^R\mathrm{d}r\, 
\frac{1}{\sqrt{|f(R)-f(r)|}}\;.
\label{dschgeod}
\end{align}

\section{Eigentime of an Inflating Observer in a Kerr-(A)dS 
	Geometry}
\label{app:geodesicKerr}
Following the same procedure for a timelike geodesic in the equatorial plane in 
Kerr-(A)dS, given by the metric (\ref{kerrads}), the normalisation of the 
four-velocity is
\begin{align}
-1=g_{tt}\,\dot t^2+g_{\phi\phi}\,\dot\phi^2+2g_{t\phi}\,\dot t\dot\phi+ 
g_{rr}\,\dot r^2\;,
\end{align}
whereas the conserved quantities induced by the Killing vector fields 
$\xi^a=\left(\frac{\partial}{\partial t}\right)^a$ and 
$\psi^a=\left(\frac{\partial}{\partial \phi}\right)^a$ read
\begin{align}
E&=-g_{ab}\xi^a u^b =-g_{tt}\,\dot t-g_{t\phi}\,\dot\phi \;, \\[2ex]
L&= g_{ab}\psi^a u^b =g_{\phi\phi}\,\dot\phi +g_{t\phi}\,\dot t \;.
\end{align}
Combining the equations and restricting again to $L=0$ yields the following 
radial equation, 
\begin{align}
\dot r^2 = \frac{E^2\,\Xi^2\left[ (r^2+a^2)^2-a^2\Delta_r 
	\right]-r^2\Delta_r}{r^4}\;.
\label{radial_equ_kerr}
\end{align}
Subsequently, we arrive at the proper distance in a Kerr-(A)dS geometry induced 
by an infalling observer in the equatorial plane, initially starting at rest at 
$r=R$ and falling towards the singularity at $r=0$:
\begin{align}
D(R)= \int_0^R\mathrm{d}r\, \frac{r^2}{\sqrt{|E^2\,\Xi^2\left[ 
		(r^2+a^2)^2-a^2\Delta_{r} \right]-r^2\Delta_{r}|}}\;,
\label{kerrgeod}
\end{align}
where $E$ is in this case then given by
\begin{align}
E^2=E^2(R)= \frac{R^2\Delta_R}{\Xi^2\left[(R^2+a^2)^2-a^2\Delta_R\right]}\;.
\end{align}

\vfill

\newpage
\bibliography{QuantumBlackHole}

\end{document}